\documentclass[a4paper]{article}
\usepackage[T1]{fontenc}
\usepackage{geometry}
\usepackage[hidelinks]{hyperref}
\usepackage{lmodern}
\usepackage{amsmath,amssymb,amsfonts,amsthm}
\usepackage[only,boxslash]{stmaryrd}
\usepackage{tikz-cd}
\usepackage{titlesec}
\usepackage[shortlabels]{enumitem}
\usepackage{calc}
\usepackage{cleveref}
\usepackage{macros}

\titleformat{\subsection}[runin]{\normalfont\bfseries}{\thesubsection}{1em}{}[.]

\theoremstyle{plain}
\theoremstyle{remark}
\newtheorem{theorem}{Theorem}
\newtheorem{lemma}[theorem]{Lemma}

\newtheorem{proposition}[theorem]{Proposition}
\newtheorem{corollary}[theorem]{Corollary}
\newtheorem{example}[theorem]{Example}
\newtheorem{remark}[theorem]{Remark}

\title{Tietze Equivalences as Weak Equivalences}
\author{Simon Henry \and Samuel Mimram}

\begin{document}
\maketitle

\begin{abstract}
  A given monoid usually admits many presentations by generators and relations and
the notion of Tietze equivalence characterizes when two presentations describe
the same monoid: it is the case when one can transform one presentation into the
other using the two families of so-called Tietze transformations. The goal of
this article is to provide an abstract and geometrical understanding of this
well-known fact, by constructing a model structure on the category of
presentations, in which two presentations are weakly equivalent when they
present the same monoid. We show that Tietze transformations form a
pseudo-generating family of trivial cofibrations and give a proof of the
completeness of these transformations by an abstract argument in this setting.

\end{abstract}


In order to navigate between the various presentations of a monoid, a very
convenient tool is provided by Tietze transformations, originally investigated
for groups~\cite{tietze1908topologischen} (see
also~\cite[chapter~II]{lyndon2015combinatorial}): these are two families of
elementary transformations one can perform on a monoid while preserving the
presented monoid. Typically, the Knuth-Bendix completion procedure for string
rewriting systems uses such transformations in order to turn a presentation of a
monoid into another presentation of the same monoid which has the property of
being convergent~\cite{knuth1983simple,guiraud2013homotopical}, and thus for
which the word problem is easily decidable. The Tietze transformations moreover
enjoy a completeness property: given any two presentations of a given monoid,
there is a way of transforming the first into the second by performing a series
of such transformations.


In this article, we provide a conceptual and geometrical point of view on Tietze
transformations, by showing that they can be abstractly thought of as
``continuously deforming'' the presentations. In order to make this formal, we
consider the category of presentations of monoids with suitably chosen morphisms
(it turns out that we need to allow some sort of degeneracies) and construct a
model structure on it, where weakly equivalent presentations are presentations
of a same monoid. We then show that the Tietze transformations can then be
interpreted in this setting as a pseudo-generating family of trivial
cofibrations: they generate trivial cofibrations with fibrant codomain. Finally,
the classical proof of completeness for Tietze transformations proceeds by
constructing some kind of cospan of Tietze transformations between two
presentations of the same monoid: we explain here how to reconstruct this proof
by purely abstract arguments based on our model structure.

The main goal of this article is thus to shed new light on theses well-known
concepts and proofs, and advocate the relevance of homotopical methods to people
working with presentations of monoids, which is why we have done our best to
have a self-contained exposition.
We see this work as a first step in order to tackle generalizations of Tietze
transformations to higher dimension (\eg coherent presentations of
categories~\cite[Section~2.1]{gaussent2015coherent}) or more involved structures
(Lawvere theories, operads, etc.). 

We recall the notion of Tietze transformation between presentations of monoids
in \cref{sec:tieze}, and of model category in \cref{sec:model}. We construct our
model structure on the category of presentations in \cref{sec:rpres-mod}, show
that Tietze transformations form a pseudo-generating family of trivial
cofibrations in \cref{sec:tietze-tcof} and use this to abstractly study Tietze
equivalences in \cref{sec:teq}.

We would like to thank Pierre-Louis Curien and Naomi Jacquet for useful
discussions.
\section{Tietze equivalences of presentations of monoids}
\label{sec:tieze}

\subsection{Monoid}

A \emph{monoid} $(M,\cdot,1)$ consists of a set~$M$ equipped with a binary
\emph{multiplication} operation $\cdot$ and a \emph{unit} element~$1$ such that
multiplication is associative and the unit acts as a neutral element. A
\emph{morphism} $f:M\to N$ between two monoids is a function which preserves
multiplication and unit. We write $\Mon$ for the resulting category.

\subsection{Free and quotient monoids}
Given a set~$X$, a \emph{word} over~$X$ is a finite sequence $u=a_1\ldots a_n$
of elements of~$X$, and its \emph{length} is $\wlen{u}=n$.
we write $X^*$ for the \emph{free monoid} generated by~$X$: its
elements are words over~$X$, multiplication $uv$ of two words is their
concatenation, and the unit is the empty word, noted~$\emptyword$.

Given a binary relation~$\sim$ on a monoid~$M$, we write $M/{\sim}$ for the
\emph{quotient monoid} whose elements and equivalence classes of elements of~$M$
by the congruence generated by~$\sim$, and multiplication and unit are induced
by those of~$M$.

\subsection{Presentation}
A \emph{presentation} $\P=\pres{\P_1}{\P_2}$ consists of
\begin{itemize}
\item a set~$\P_1$ of \emph{generators},
\item a set $\P_2\subseteq\freemon\P_1\times\freemon\P_1$ of \emph{relations}.
\end{itemize}
Such a presentation is \emph{finite} when both the sets~$\P_1$ and~$\P_2$ are. A
relation $(u,v)\in\freemon\P_1$ is generally denoted by ``$u\rto v$'' and we
write $\psim\P$ for the smallest congruence generated by~$\P_2$. A
\emph{morphism} $f:\P\to\Q$ between presentations is a function $f:\P_1\to\Q_1$
such that, for every $u\rto v\in\P_2$, we have $f^*(u)\rto f^*(v)\in\Q_2$. A
\emph{subpresentation} $\P'$ of~$\P$ is a presentation equipped with a morphism
$\P'\to\P$ whose underlying function is an inclusion.
We write $\Pres$ for the category of presentations and their morphisms. Note
that, by definition, there is a forgetful functor $\Pres\to\Set$ sending a
presentation~$\P$ to its set~$\P_1$ of generators.

\subsection{Presented monoid}
\label{sec:pmon}
The monoid $\pmon\P$ \emph{presented} by a presentation~$\P$ is the quotient
monoid $\pmon\P=\freemon\P_1/\P_2$ \ie the quotient of the free monoid
$\freemon\P_1$ by the congruence $\psim\P$ generated by~$\P_2$. We often write
$q^\P:\freemon\P_1\to\pmon\P$ for the quotient morphism and, given
$u\in\freemon\P_1$, we write $\pmon u$ for its equivalence class $q^\P(u)$. More
generally, we say that a monoid~$M$ is \emph{presented} by~$\P$ when~$M$ is
isomorphic to~$\pmon\P$, what we sometimes write $M\isoto\pres{\P_1}{\P_2}$.
This construction extends as a functor $\Pres\to\Mon$.

\begin{example}
  We have the following presentations:
  \begin{align*}
    \N&\isoto\pres{a}{}
    &
    \N\times\N&\isoto\pres{a,b}{ab\rto ba}
    \\
    \N/2\N&\isoto\pres{a}{aa\rto 1}
    &
    \Z&\isoto\pres{a,b}{ab\rto 1,ba\rto 1}
    \pbox.
  \end{align*}
\end{example}

\subsection{Standard presentation}
To any monoid~$M$, one can associate a presentation~$\stdpres M$, called the
\emph{standard presentation} of~$M$, defined by
\begin{align*}
  \P_1&=\setof{\ul a}{a\in M}\\
  \P_2&=\setof{\ul a_1\ldots \ul a_n\rto\ul b_1\ldots \ul b_m}{a_1\ldots a_n=b_1\ldots b_m}
  \pbox,
\end{align*}
\ie it contains the elements of the monoids as generators and there is a
relation between two words of generators when the product of their elements are
equal. This construction extends as a functor $\Mon\to\Pres$. It can be used to
show that any monoid admits at least one presentation:

\begin{lemma}
  Given a monoid~$M$, its standard presentation is a presentation of~$M$:
  $\pmon{\stdpres M}\isoto M$.
\end{lemma}
  % ...

\begin{lemma}
  \label{lem:pmon-adj}
  The presentation functor is left adjoint to the standard presentation functor
   \[
    \begin{tikzcd}
      \Pres\ar[r,bend left,"\pmon-"]\ar[r,phantom,"\bot"]&\ar[l,bend left,"\stdpres-"]\Mon
    \end{tikzcd}
  \]
  the counit of the adjunction being an isomorphism.
\end{lemma}

\subsection{Reflexive presentations}
\label{sec:refl-pres}
A presentation $\P$ is \emph{reflexive} when for every word $u\in\freemon\P_1$
there is a relation $u\rto u\in\P_2$. We write $\rPres$ for the full subcategory
of $\Pres$ on reflexive presentations.

\begin{lemma}
  The expected forgetful functor admits a left adjoint
  \[
    \begin{tikzcd}
      \Pres\ar[r,bend left]\ar[r,phantom,"\bot"]&\ar[l,bend left]\rPres
    \end{tikzcd}
  \]
  sending a presentation $\P$ to the presentation $\Q$ with
  \begin{align*}
    \Q_1&=\P_1
    &
    \Q_2=\P_2\cup\setof{u\rto u}{u\in\freemon\P_1}
  \end{align*}
  and $\rPres$ is equivalent to the Kleisli category of the monad on $\Pres$
  induced by the adjunction.
\end{lemma}

\begin{lemma}
  The category $\rPres$ is equivalent to the category whose objects are
  presentations (not necessarily reflexive) and a morphism $f:\P\to\Q$ is a
  function $f:\P_1\to\Q_1$ such that for every relation $u\rto v\in\P_2$ we have
  either $f(u)\rto f(v)\in\Q_2$ or $f(u)=f(v)$.
\end{lemma}

\noindent
In the following, when describing concrete examples of reflexive presentations,
we generally omit mentioning reflexivity relations (or, alternatively, the
description of morphisms given by previous lemma could be considered).

\begin{remark}
  \label{rem:refl-stdpres}
  The standard presentation is clearly reflexive and thus the adjunction of
  \lemr{pmon-adj} restricts to an adjunction between reflexive presentations and
  monoids.
\end{remark}

\subsection{Equivalence between presentations}
There is a very natural notion of equivalence of presentations: two
presentations can be considered as \emph{equivalent} when they present
isomorphic monoids. In order to provide a concrete and amenable description of
this relation, Tietze has introduced a family of transformations on
presentations which characterize the equivalence. Those were originally
formulated in the context of presentations of
groups~\cite{tietze1908topologischen}.

We begin with a simpler but useful characterization of the equivalence:

\begin{lemma}
  \label{lem:eq-cospan}
  Two presentations~$\P$ and~$\Q$ are such that $\pmon\P\isoto\pmon\Q$ if and
  only if there is a cospan of presentations
  \[
    \begin{tikzcd}
      \P\ar[r,"f"]&\R&\ar[l,"g"']\Q
    \end{tikzcd}
  \]
  such that the induced monoid morphisms $\pmon f:\pmon\P\to\pmon\R$ and
  $\pmon g:\pmon\Q\to\pmon\R$ are isomorphisms.
\end{lemma}
\begin{proof}
  If there is a cospan as above then we have $\pmon\P\isoto\pmon\R\isoto\pmon\Q$
  and~$\P$ and~$\Q$ are thus equivalent. Conversely, suppose that~$\P$ presents
  the monoid~$M$, \ie there is an isomorphism $\pmon\P\to M$. Under the
  adjunction of \lemr{pmon-adj}, this induces a map $f:\P\to\stdpres M$ such
  that $\pmon f:\pmon\P\to\pmon{\stdpres M}=M$. Similarly, we can construct a
  map $g:\P\to\stdpres M$.
\end{proof}

\subsection{Tietze transformation}
The \emph{elementary Tietze transformations} are the following transformations
producing a new presentation~$\Q$ from a presentation~$\P$:
\begin{enumerate}[labelwidth=\widthof{\trel},leftmargin=!]
\item[\tgen] \emph{adding a derivable generator}: given a new generator
  $a\not\in\P_1$ and word $u\in\freemon{\P_1}$, we define the presentation~$\Q$
  by
  \begin{align*}
    \Q_1&=\P_1\sqcup\set{a}
    &
    \Q_2&=\P_2\cup\set{u\rto a}
    \pbox,
  \end{align*}
\item[\trel] \emph{adding a derivable relation}: given two words
  $u,v\in\freemon\P_1$ such that $u\psim\P v$, we define the presentation~$Q$ by
  \begin{align*}
    \Q_1&=\P_1\phantom{\sqcup\set{a}}
    &
    \Q_2&=\P_2\cup\set{u\rto v}
    \pbox.
  \end{align*}
\end{enumerate}
It is easy to see that those transformations preserve the presented monoids:

\begin{lemma}
  \label{lem:tietze-iso}
  Given an elementary Tietze transformation from~$\P$ to~$\Q$, we have an
  isomorphism $\pmon\P\isoto\pmon\Q$.
\end{lemma}

\noindent
A \emph{Tietze transformation} from $\P$ to~$\Q$ consists in a finite sequence
of presentations
\[
  \P=\P^0, \P^1, \P^2, \ldots, \P^n=\Q
\]
such that for every $i$ with $0\leq i<n$ there is an elementary Tietze
transformation from $\P^i$ to~$\P^{i+1}$. In this situation, we sometimes write
\[
  \P\tto\Q
\]
Note that contrarily to the usual convention, we do not allow here removing
generators or relations.

The transformation \trel{} can be replaced by the following four transformations:
\begin{enumerate}[labelwidth=\widthof{\tsym},leftmargin=!]
\item[\trefl] \emph{reflexivity}: given $u\in\freemon\P_1$, we define~$\Q$ by
  \begin{align*}
    \Q_1&=\P_1
    &
    \Q_2&=\P_2\cup\set{u\rto u}
    \pbox,
  \end{align*}
\item[\tsym] \emph{symmetry}: given $u,v\in\freemon\P_1$ such that
  $u\rto v\in\P_2$, we define~$\Q$ by
  \begin{align*}
    \Q_1&=\P_1
    &
    \Q_2&=\P_2\cup\set{v\rto u}
    \pbox,
  \end{align*}
\item[\ttrans] \emph{transitivity}: given $u,v,w\in\freemon\P_1$ such that
  $u\rto v,v\rto w\in\P_2$, we define~$\Q$ by
  \begin{align*}
    \Q_1&=\P_1
    &
    \Q_2&=\P_2\cup\set{u\rto w}
    \pbox.
  \end{align*}
\item[\tctxt] \emph{context}: given $u,v,v',w\in\freemon\P_1$ such that
  $v\rto v'\in\P_2$, we define~$\Q$ by
  \begin{align*}
    \Q_1&=\P_1
    &
    \Q_2&=\P_2\cup\set{uvw\rto uv'w}
    \pbox,
  \end{align*}
\end{enumerate}
The resulting systems are the same in the following sense:

\begin{lemma}
  \label{lem:alt-tietze}
  The following assertions are equivalent: there is a Tietze transformation
  from~$\P$ to~$\Q$
  \begin{enumerate}[(i)]
  \item using \tgen{} and \trel{},
  \item using \tgen{}, \trefl{}, \tsym{}, \ttrans{} and \tctxt{}.
  \end{enumerate}
\end{lemma}

\noindent
In the following, unless otherwise mentioned, we use the second set of Tietze
transformations which are easier to work with because they are more ``atomic''.

\subsection{Tietze equivalence}
A \emph{Tietze equivalence} from $\P$ to~$\Q$ is a finite sequence of
presentations $\P=\P^0,\P^1,\P^2,\ldots,\P^n=\Q$ such that for every $i$ with
$0\leq i<n$ there is a Tietze transformation from $\P^i$ to $\P^{i+1}$ or from
$\P^{i+1}$ to~$P^{i}$.
Two presentations are \emph{Tietze equivalent} when there is a Tietze
equivalence between them. Otherwise said, the Tietze equivalence is the smallest
equivalence relation relating any two presentations between which there is an
(elementary) Tietze transformation.
By \lemr{tietze-iso} above, Tietze equivalences preserve the presented
monoids. It well known that, for finite presentations, the converse
holds~\cite[chapter II]{lyndon2015combinatorial}:

\begin{theorem}
  \label{thm:tietze}
  Given two finite presentations $\P$ and $\Q$, we have $\pmon\P\isoto\pmon\Q$
  if and only if $\P$ and $\Q$ are Tietze equivalent.
\end{theorem}
\begin{proof}
  The right-to-left implication follows from \lemr{tietze-iso}.
  For the left-to-right implication, suppose given an isomorphism
  $\pmon\P\isoto\pmon\Q$. For the sake of simplicity we suppose that we actually
  have $\pmon\P=\pmon\Q$ and more generally that Tietze equivalent presentations give rise
  to identical presented monoids (the proof without this assumption can be
  constructed from the one below by inserting isomorphisms at required
  places).
  Given a generator $a\in\P_1$, there exists an element~$u\in\freemon\Q_1$ such
  that $q^\P(a)=q^\Q(u)$. We write $a^\Q$ for a choice of such an
  element. Dually, given $b\in\Q_1$, we write $b^\P\in\freemon\P_1$ for a word
  such that $q^\P(b^\P)=q^\Q(b)$.
  We generalize this notation to words $u=a_1\ldots a_n\in\freemon\P_1$, by
  setting $u^\Q=a_1^\Q\ldots b_n^\Q$ (and we define $v^\Q$ for
  $v\in\freemon\Q_1$ similarly). Note that, for $u\in\freemon\P_1$, we have
  \begin{equation}
    \label{eq:uqp}
    (u^\Q)^\P\psim\P u    
  \end{equation}
  (and dually).
  We construct a presentation~$\R$ by
  \begin{align*}
    \R_1&=\P_1\sqcup\Q_1
    &
    \R_2&=\P_2\sqcup\Q_2\sqcup\R_2^\P\sqcup\R_2^\Q
  \end{align*}
  where
  \begin{align*}
    \R_2^\P&=\setof{a^\Q\rto a}{a\in\P_1}
    &
    \R_2^\Q&=\setof{b^\P\rto b}{b\in\Q_1}
  \end{align*}
  We now construct a Tietze transformation from~$\P$ to~$\R$. Dually, we will be
  able to construct a Transformation from~$\Q$ to~$\R$ and we will be able to
  conclude that $\P$ and~$\Q$ are Tietze equivalent:
  \[
    \P\tto\R\tot\Q\pbox.
  \]
  By using Tietze transformations \tgen{}, starting from~$\P$, we can add each
  generator $b\in\Q_1$ along with the relation $b^\P\rto b$, thus obtaining a
  transformation
  \[
    \P=\pres{\P_1}{\P_2}\tto\P'=\pres{\P_1,\Q_1}{\P_2,\R_2^\Q}\pbox.
  \]
  Note that, given a word $u\in\freemon\Q_1$, we have
  $u^\P\psim{\P'}u$. Therefore, given $a\in\P_1$, we have
  $a^\Q\psim{\P'}(a^\Q)^\P\psim{\P'}a$ by \eqref{eq:uqp}. By using Tietze
  transformations \trel{} we can add each derivable relation $a^\Q\rto a$
  to~$\P'$ thus reaching the presentation~$\R$:
  \[
    \P\tto\P'\tto\R
    \pbox.\qedhere
  \]
\end{proof}

\begin{remark}
  The proof above uses Tietze transformations \tgen{} and \trel{}. The proof can
  be performed by using the other set of transformations given by
  \lemr{alt-tietze}, at the cost of having to take a slightly bigger~$\R$.
\end{remark}

\noindent
The proof of \thmr{tietze} constructs a ``cospan'' of Tietze transformations. We
will see that it can be constructed by using tools coming from model categories.

\subsection{An example}
\label{sec:ex-tietze}
Consider the presentations
\[
  \pres{a}{}
  \qquad\text{and}\qquad
  \pres{a,b}{b\rto bb,1\rto bb}
  \pbox.
\]
Both present the additive monoid~$\N$, and indeed there is a Tietze equivalence
between them:
\begin{align*}
  \pres{a}{}
  &\to\pres{a,b}{1\rto b}&&\text{\tgen}\\
  &\to\pres{a,b}{1\rto b,b\rto bb}&&\text{\tctxt}\\
  &\to\pres{a,b}{1\rto b,b\rto bb,1\rto bb}&&\text{\ttrans}\\
  &\to\pres{a,b}{1\rto b,b\rto bb,1\rto bb,bb\rto b}&&\text{\tsym}\\
  &\ot\pres{a,b}{b\rto bb,1\rto bb,bb\rto b}&&\text{\ttrans}\\
  &\ot\pres{a,b}{b\rto bb,1\rto bb}&&\text{\tsym}
\end{align*}
Also note that both presentations are ``minimal'': there is no way to remove a
derivable generator or a relation without changing the presented monoid. In
particular, starting from the second presentation, we have to add relations
first in order to be able remove the generator~$b$ and all the relations.

\subsection{Generalization to infinite presentations}
The above \cref{thm:tietze} holds only for finite presentations, which is the
way it is usually stated. It can easily be generalized to presentations of
arbitrary cardinality by allowing the Tietze transformations to add \emph{sets}
of derivable generators and \emph{sets} of derivable relations (instead of only
one), what we call \emph{generalized Tietze transformations}. The right way to
think of those is as being obtained as cellular extensions of elementary Tietze
transformations and we will prove in \thmr{gen-tietze} the following
generalization of \cref{thm:tietze}, which was already known, see for
instance~\cite[section~1.5]{magnus2004combinatorial}:

\begin{theorem}
  Given two presentation $\P$ and~$\Q$, we have $\pmon\P\isoto\pmon\Q$ if and
  only if they are related by a zig-zag of generalized Tietze transformations,
  \ie there exists a finite sequence of presentations
  \[
    \P=\P^0,\P^1,\ldots,\P^{2n}=\Q
  \]
  such that for every index~$i$, there is a generalized Tietze transformation
  from $\P^{2i}$ to $\P^{2i+1}$ and from $P^{2i+2}$ to $\P^{2i+1}$.
\end{theorem}

\begin{remark}
  The naive generalization of \cref{thm:tietze}, which states that two
  presentations have the same presented monoid if and only if they are related
  by a ``possibly infinite zig-zag'' of elementary Tietze transformations, is
  plain wrong (and this is not what the above theorem states). For instance,
  consider the following presentation of the monoid~$\N$:
  \[
    \P=\pres{a,b_i}{a\To b_i,b_i\To b_{i+1}}_{i\in\N}
  \]
  and write $\P^i$ for $\P$ with the relations $a\To b_i$ removed for $i<k$. We
  have $\P^0=\P$ and the relation $a\To b_k$ is derivable in~$\P^k$, so that
  there is an elementary Tietze transformation from $\P^{k+1}$ to
  $\P^k$. However, writing
  \[
    \P^\infty=\pres{a,b_i}{b_i\To b_{i+1}}_{i\in\N}
  \]
  we have that $\P^0$ does not present the same monoid as $\P^\infty$ even
  though there is an ``infinite sequence of elementary Tietze transformations''
  between them. Namely, $\P$ presents $\N$ whereas $\P^\infty$ presents $\N*\N$,
  the free product of two copies of~$\N$, and two are not isomorphic (the former
  is commutative whereas the later is not).
\end{remark}
\section{Model categories}
\label{sec:model}
In this section, we recall elementary definitions and facts about model
categories which we will use in the following and refer the reader to classical
textbooks for details~\cite{hovey2007model}.

\subsection{Lifting properties}
Suppose fixed a category. A morphism $p:X\to Y$ has the \emph{right lifting
  property}, or \emph{rlp}, with respect to a morphism $i:A\to B$ when for every
pair of morphisms $f:A\to X$ and $g:B\to Y$ such that $p\circ f=g\circ i$ there exists a
morphism $h:B\to X$ making the following diagram commute:
\[
  \begin{tikzcd}
    A\ar[d,"i"']\ar[r,"f"]&X\ar[d,"p"]\\
    B\ar[r,"g"']\ar[ur,dotted,"h"description]&Y
  \end{tikzcd}
\]
In this situation, we also say that $i$ has the \emph{left lifting property}, or
\emph{llp}, with respect to~$f$, and write $i\boxslash p$. Given two
classes~$\clL$ and $\clR$, we write $\clL\boxslash\clR$ whenever $i\boxslash p$
for every $i\in\clL$ and $p\in\clR$. We also write $\rlp{\clL}$ (\resp
$\llp{\clR}$) for the class of morphism with the rlp (\resp llp) with respect
to~$\clL$ (\resp $\clR$).

\begin{lemma}
  \label{lem:ortho}
  Given classes of morphisms~$\clL$, $\clL'$, $\clR$ and $\clR'$,
  \begin{align*}
    \clL&\subseteq\lrlp\clL
    &
    \rlp{(\lrlp\clL)}&=\rlp\clL
    &
    \clL\subseteq\clL'&\text{ implies }\rlp\clL\supseteq\rlp{\clL'}
    \pbox,
    \\
    \clR&\subseteq\rllp\clR
    &
    \llp{(\rllp\clR)}&=\llp\clR
    &
    \clR\subseteq\clR'&\text{ implies }\llp\clR\supseteq\llp{\clR'}
    \pbox.
  \end{align*}
\end{lemma}

\begin{lemma}
  \label{lem:llp-stab}
  We suppose the ambient category cocomplete. A class of the form
  $\clL=\llp\clR$ contains isomorphisms and is closed under
  \begin{itemize}
  \item coproducts: for any family $(i_k:A_k\to B_k)_{k\in K}$ of morphisms
    in~$\clL$, the morphism
    \[
      \coprod_{k\in K}i_k:\coprod_{k\in K}A_k\to\coprod_{k\in K}B_k
    \]
    is also in the class,
  \item pushouts: for any morphism $i:A\to B$ in~$\clL$ and morphism $f:A\to A'$,
    for any pushout diagram
    \[
      \begin{tikzcd}
        A\ar[d,"i"']\ar[r,"f"]\ar[phantom,dr,"\ulcorner",very near end]&A'\ar[d,"j"]\\
        B\ar[r]&B'
      \end{tikzcd}
    \]
    the morphism $j$ also belongs to $\clL$,
  \item countable compositions: for any diagram
    \begin{equation}
      \label{eq:ccomp}
      \begin{tikzcd}
        A_0\ar[r,"f_0"]&A_1\ar[r,"f_1"]&\cdots
      \end{tikzcd}
    \end{equation}
    consisting of morphisms $f_k:A_k\to A_{k+1}$ in $\clL$ for $k\in\N$, the
    canonical morphism
    \[
      A_0\to\colim_k A_k
    \]
    also belongs to~$\clL$,
  \item retracts:
    given a morphism $i:A\to B$ and two retracts $r\circ s=\id_{A'}$ and
    $r'\circ s'=\id_{B'}$, any morphism $j:A'\to B'$ for which there is a
    commutative diagram
    \begin{equation}
      \label{eq:retract}
      \begin{tikzcd}
        A'\ar[rr,bend left,"\id_{A'}"]\ar[d,"j"']\ar[r,"s"']&A\ar[d,"i"description]\ar[r,"r"']&A'\ar[d,"j"]\\
        B'\ar[rr,bend right,"\id_{B'}"']\ar[r,"s'"]&B\ar[r,"r'"]&B'
      \end{tikzcd}
    \end{equation}
    also belongs to~$\clL$.
  \end{itemize}
  \label{lem:rlp-stab}
  Dually, any class for the form $\rlp\clL$ contains isomorphisms and is closed
  under products, pullbacks, countable compositions and retracts.
\end{lemma}

\noindent
Given a class $\clI$ of morphisms, the class $\cell\clI$ of
\emph{$\clI$-cellular extensions} is defined as the smallest class of morphisms
closed under sums, pushouts and countable compositions (note that we do not
require closure under retracts).

\begin{lemma}
  A morphism is an $\clI$-cellular extension if and only if it is a composite of
  pushouts of sums of elements of~$\clI$.
\end{lemma}

\begin{lemma}
  \label{lem:cell-cof}
  Given a class $\clI$ of morphisms, the class of $\clI$-cellular
  extensions is included in $\lrlp\clI$.
\end{lemma}
\begin{proof}
  By \lemr{ortho}, we have $\clI$ included in $\lrlp\clI$ and, by
  \lemr{llp-stab}, this class is closed under sums, pushouts and countable
  compositions.
\end{proof}

\begin{lemma}[Retract lemma]
  \label{lem:retract}
  Given a factorization $f=p\circ i$ such that $f\boxslash p$, $f$ is a retract
  of~$i$.
  Dually, given a factorization $f=p\circ i$ such that $i\boxslash f$,
  $f$ is a retract of~$p$.
\end{lemma}
\begin{proof}
  Since $f\boxslash p$, we have a map $h$ such that
  \[
    \begin{tikzcd}
      X\ar[d,"f"']\ar[r,"i"]&Y\ar[d,"p"]\\
      Z\ar[r,equal]\ar[ur,"h"description]&Z
    \end{tikzcd}
  \]
  and the map~$f$ is thus a retract of~$i$:
  \[
    \begin{tikzcd}
      X\ar[d,,"f"']\ar[r,equal]&X\ar[d,"i"description]\ar[r,equal]&X\ar[d,"f"]\\
      Z\ar[r,"h"']&Y\ar[r,"p"']&Z
    \end{tikzcd}
  \]
  as claimed.
\end{proof}

\subsection{Weak factorization system}
A \emph{weak factorization system} on a category is a pair $(\clL,\clR)$ of
classes of morphisms such that
\begin{itemize}
\item every morphism $f$ factors as $f=p\circ i$ with $i\in\clL$ and $p\in\clR$,
\item $\clL=\llp\clR$ and $\clR=\rlp\clL$.
\end{itemize}

\begin{remark}
  \label{rem:wfs-retract}
  From~\lemr{llp-stab} and \lemr{retract}, the second condition can be
  equivalently be replaced by the two following conditions
  \begin{itemize}
  \item $\clL\boxslash\clR$,
  \item the classes $\clL$ and $\clR$ are closed under retracts.
  \end{itemize}
\end{remark}

\noindent
One of the main techniques in order to construct weak factorization systems is
due to the following proposition~\cite[Section~2.1.2]{hovey2007model}. The
notion of locally finitely presentable category is recalled in \secr{loc-pres}.

\begin{proposition}[Small object argument]
  \label{prop:small-obj}
  Suppose that the category is cocomplete and locally finitely presentable. For
  any class~$\clI$ of morphisms, $(\lrlp\clI,\rlp\clI)$ is a weak factorization
  system. Moreover, every morphism~$f$ factors as $f=p\circ i$ where
  $i\in\lrlp\clI$ is an $\clI$-cellular extension and $p\in\rlp\clI$. Moreover,
  every element of $\lrlp\clI$ is a retract of an $\clI$-cellular extension.
\end{proposition}

\subsection{Model category}
A \emph{model category} is a category equipped with three classes of morphisms
\begin{itemize}
\item $\clC$: cofibrations,
\item $\clW$: weak equivalences,
\item $\clF$: fibrations
\end{itemize}
such that
\begin{itemize}
\item the category is complete and cocomplete,
\item weak equivalences satisfy the 2-out-of-3 property: given a diagram
  \[
    \begin{tikzcd}
      &\ar[dr,"g"]\\
      \ar[ur,"f"]\ar[rr,"g\circ f"']&&{}
    \end{tikzcd}
  \]
  if two morphisms belong to~$\clW$ then so does the third,
\item $(\clC,\clW\cap\clF)$ forms a weak factorization system,
\item $(\clC\cap\clW,\clF)$ forms a weak factorization system.
\end{itemize}
An object $X$ is \emph{cofibrant} when the initial morphism $\iobj\to X$ is a
cofibration, and \emph{fibrant} when the terminal morphism $X\to\tobj$ is a
fibration.





From the previous section, we can expect that the weak factorization system can be
generated as lifting completions of some classes. Indeed, many model categories
are \emph{cofibrantly generated} (also sometimes called \emph{combinatorial}
since we work here with locally presentable
categories)~\cite[Theorem~2.1.19]{hovey2007model}:

\begin{proposition}
  \label{prop:cof-gen}
  In a locally presentable, complete and cocomplete category, suppose given a
  subcategory~$\clW$ closed under retracts and satisfying the 2-out-of-3
  property, and two sets $\clI$ and $\clJ$ of morphisms such that the inclusions
  \begin{align*}
    \rlp\clI&\subseteq\rlp\clJ\cap\clW
    &
    \lrlp\clJ&\subseteq\lrlp\clI\cap\clW
  \end{align*}
  hold, one of them being an equality. Then we have a model category with $\clW$
  as weak equivalences, $\lrlp\clI$ as cofibrations and $\rlp\clJ$ as
  fibrations. In this case, the elements of $\clI$ as $\clJ$ are respectively
  called \emph{generating cofibrations} and \emph{generating trivial
    cofibrations}.
\end{proposition}

\section{A model structure on reflexive presentations}
\label{sec:rpres-mod}
Our aim is to construct a model structure on the category of reflexive
presentations where weak equivalences correspond to presenting isomorphic
categories and trivial cofibrations are Tietze transformations. The general
strategy here is to use \propr{cof-gen} and thus to satisfy all the required
hypotheses: in particular, we want to show the equality
$\rlp\clI=\rlp\clJ\cap\clW$. Unless otherwise mentioned, all the presentations
considered in this section are supposed to be reflexive; the reason for this
shall be discussed in \secr{nr-pres}.
We first study some of the properties of the category of reflexive
presentations.

\subsection{Colimits}
\label{sec:pres-colim}
The category~$\rPres$ has coproducts. Namely, given two presentations~$\P$
and~$\Q$, their coproduct $\P\sqcup\Q$ is given by
\begin{align*}
  (\P\sqcup\Q)_1&=\P_1\sqcup\Q_1
  &
  (\P\sqcup\Q)_2&=\P_2\sqcup\Q_2
\end{align*}
and the argument generalizes to show that the category has small coproducts. In
particular, the initial object~$\iobj$ is the empty presentation, with
$\iobj_1=\emptyset$ and $\iobj_2=\emptyset$. Suppose given two morphisms of
presentations
\[
  \begin{tikzcd}
    \P\ar[r,"f",shift left]\ar[r,"g"',shift right]&\Q
  \end{tikzcd}
\]
Their coequalizer is the presentation~$\R$ whose set of generators is the
coequalizer
\[
  \begin{tikzcd}
    \P_1\ar[r,"f",shift left]\ar[r,"g"',shift right]&\Q_1\ar[r,dotted,"h"]&\R_1
  \end{tikzcd}
\]
\ie the quotient set $\R_1=\Q_1/{\sim}$ under the smallest equivalence relation
such that $f(a)\sim f(b)$ for $a\in\P_1$, the function $h$ being the quotient
map, and the set of relations is
\[
  \R_2=\setof{\freemon h(u)\rto\freemon h(v)}{u\rto v\in\Q_1}
  \pbox.
\]
The category is thus cocomplete. In particular, the pushout of a diagram
\[
  \begin{tikzcd}
    \Q^1&\ar[l,"f^1"']\P\ar[r,"f^2"]&\Q^2
  \end{tikzcd}
\]
is the presentation~$\R$ whose set~$\R_1$ of generators is the pushout of the
underlying sets of generators, with cocoone maps $h^1:\Q^1\to\R$ and
$h^2:\Q^2\to\R$, and relations
\[
  R_2=
  \setof{h^1(u)\rto h^1(v)}{u\rto v\in\Q^1_2}
  \cup
  \setof{h^2(u)\rto h^2(v)}{u\rto v\in\Q^2_2}
  \pbox.
\]
Note that the forgetful functor $\rPres\to\Set$, sending a presentation to its
underlying set of generators, preserves colimits.

\subsection{Limits}
\label{sec:pres-lim}
The product $\P\times\Q$ of two reflexive presentations~$\P$ and~$\Q$ has
generators $(\P\times\Q)_1=\P_1\times\Q_1$ and the set $(\P\times\Q)_2$ of
relations is
\[
  \setof{(a_1,a'_1)\ldots(a_m,a'_m)\rto(b_1,b'_1)\ldots(b_n,b'_n)}{
    \begin{array}{r@{\ }l}
      a_1\ldots a_m\rto b_1\ldots b_n&\in\P_2\\
      a'_1\ldots a'_m\rto b'_1\ldots b'_n&\in\Q_2
    \end{array}
  }
  \pbox.
\]
This generalizes to small products. In particular, the terminal
presentation~$\tobj$ has one generator~$a$ and all relations of the form
$a^m\rto a^n$ for $m,n\in\N$. Given two morphisms of presentations
\[
  \begin{tikzcd}
    \P\ar[r,"f",shift left]\ar[r,"g"',shift right]&\Q
  \end{tikzcd}
\]
their equalizer~$\R$ is given by
\[
  \R_1=\setof{a\in\P_1}{f(a)=g(a)}
\]
\ie this is the equalizer of the underlying sets, and relations are
\[
  \R_2=\setof{u\rto v\in\P_2}{\freemon f(u)=\freemon g(u)\text{ and }\freemon f(v)=\freemon g(v)}
  \pbox.
\]
The category is thus complete and the forgetful functor $\rPres\to\Set$
preserves limits.

\subsection{Monomorphisms}
A monomorphism $f:\P\to\Q$ is a morphism whose underlying function
$f:\P_1\to\Q_1$ is injective, \ie the forgetful functor $\rPres\to\Set$ reflects
monomorphisms. In this sense, the monomorphisms of presentations inherit the
properties of those of the categories of sets. For instance,

\begin{lemma}
  \label{lem:mono-stab}
  In~$\rPres$, monomorphisms are stable under coproducts, pushouts and countable
  compositions.
\end{lemma}
\begin{proof}
  The forgetful functor to sets preserves coproducts, pushouts and countable
  compositions, and reflects monomorphisms.
\end{proof}

\begin{remark}
  These stability conditions are not generally true in a category. As a
  counter-example, in the category of commutative rings, the inclusion
  $i:\Z\to\mathbb{Q}$ is a mono, but the sum (which is here the tensor product,
  and corresponds to the usual tensor product of $\Z$-modules)
  \[
    \id_{\Z/2}\otimes i:\Z/2=\Z/2\otimes\Z\to\Z/2\otimes\mathbb{Q}=1
  \]
  is not a mono. It is however the case that monomorphisms are stable under
  pushout in a topos (and, more generally, an adhesive category).
\end{remark}

\subsection{Epimorphisms}
\label{sec:pres-epi}
Similarly, an epimorphism $f:\P\to\Q$ is a morphism whose underlying
function~$\P_1\to\Q_1$ is a surjection.

\subsection{Local finite presentability}
\label{sec:loc-pres}
We refer to~\cite{adamek1994locally} for a detailed presentations of the notions
introduced here. An object~$X$ of a category~$\C$ is \emph{finitely
  presentable} when the representable functor
\[
  \Hom(X,-):\C\to\Set
\]
preserves filtered limits: this means that for a diagram $(Y_i)_{i\in I}$
indexed by a filtered category~$I$, the canonical morphism
\[
  \colim_i\Hom(X,Y_i)\to\Hom(X,\colim_iY_i)
\]
is an isomorphism. In particular, finitely presentable presentations objects are
precisely the finite presentations.

A locally small category~$\C$ is \emph{locally finitely presentable} when it is
cocomplete and there is a set of finitely presentable objects such that every
object of~$\C$ is a filtered colimit of objects in this set. In the case of the
category of presentations, every presentation is the filtered colimit of its
finite subpresentations, and the category~$\rPres$ is thus locally finitely
presentable. The category $\Mon$ is also locally finitely presentable, as the
category of models of a Lawvere theory.


\subsection{Weak equivalences}
\label{sec:pres-w}
We write $\clW$ for the class of morphisms $f:\P\to\Q$ such that the induced
morphism $\pmon f:\pmon\P\to\pmon\Q$ between presented monoids is an
isomorphism. Many of the properties of isomorphisms are thus reflected on weak
equivalences:

\begin{lemma}
  \label{lem:pres-w-stab}
  \label{lem:pres-w-23}
  \label{lem:pres-w-comp}
  \label{lem:pres-w-retract}
  The class $\clW$ satisfies the 2-out-of-3 property and is closed under
  coproducts, pushouts, countable compositions and retracts.
\end{lemma}
\begin{proof}
  The class of isomorphisms in any category satisfies the 2-out-of-3
  property. Isomorphisms are closed under sums. Namely, given two isomorphisms
  $i:A\to B$ and $i':A\to B$, the two following diagrams commute:
  
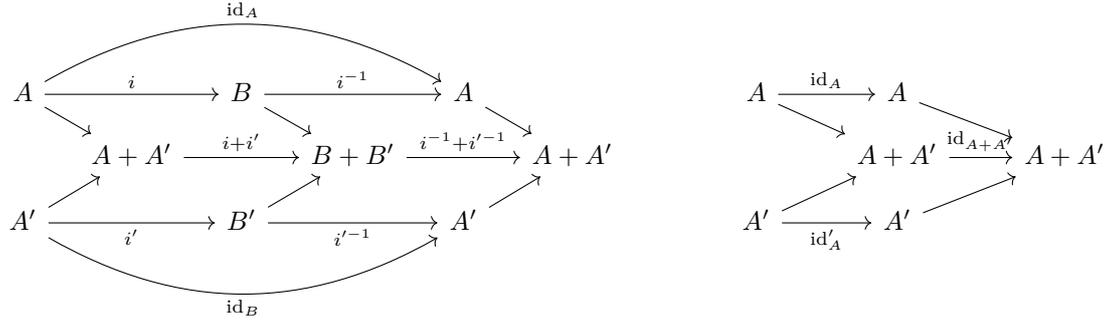
\[
    \begin{tikzcd}[sep=small]
      A\ar[rr,"i"]\ar[dr]\ar[rrrr,bend left,"\id_A"]&&B\ar[dr]\ar[rr,"i^{-1}"]&&A\ar[dr]\\
      &A+A'\ar[rr,"i+i'"]&&B+B'\ar[rr,"i^{-1}+i'^{-1}"]&&A+A'\\
      A'\ar[ur]\ar[rr,"i'"']\ar[rrrr,bend right,"\id_B"']&&B'\ar[ur]\ar[rr,"i'^{-1}"']&&A'\ar[ur]
    \end{tikzcd}
    \qquad\qquad
    \begin{tikzcd}[row sep=small]
      A\ar[dr]\ar[r,"\id_A"]&A\ar[dr]\\
      &A+A'\ar[r,"\id_{A+A'}"]&A+A'\\
      A'\ar[ur]\ar[r,"\id_A'"']&A'\ar[ur]
    \end{tikzcd}
  \]
  By universal property of coproducts, we deduced that
  $(i^{-1}+i'^{-1})\circ(i+i')=\id_{A+A'}$. Similarly, we can show
  $(i+i')\circ(i^{-1}+i'^{-1})=\id_{B+B'}$ and $i+i'$ is thus an isomorphism.
  Isomorphisms are also closed under pushouts. Namely, consider $j:A'\to B'$ the
  pushout of an isomorphism $i:A\to B$ along a morphism $f:A\to A'$. The two
  following diagrams commute, where $j':B'\to A'$ is defined by universal
  property of the pushout:
  
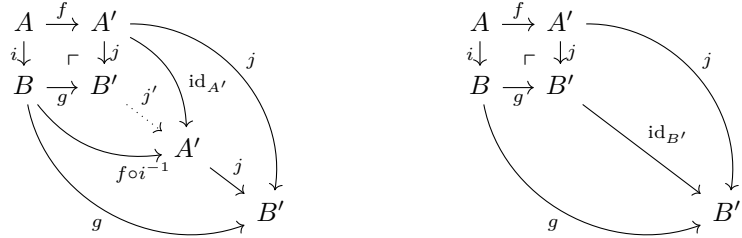
\[
    \begin{tikzcd}[sep=small]
      A\ar[d,"i"']\ar[r,"f"]\ar[phantom,dr,"\ulcorner",very near end]&A'\ar[d,"j"]\ar[ddr,bend left,"\id_{A'}",near end]\ar[dddrr,bend left=50,"j"]\\
      B\ar[r,"g"']\ar[drr,bend right,"f\circ i^{-1}"',very near end]\ar[ddrrr,bend right=50,"g"']&B'\ar[dr,dotted,"j'",near start]\\
      &&A'\ar[dr,"j"]\\
      &&&B'
    \end{tikzcd}
    \qquad\qquad\qquad
    \begin{tikzcd}[sep=small]
      A\ar[d,"i"']\ar[r,"f"]\ar[phantom,dr,"\ulcorner",very near end]&A'\ar[d,"j"]\ar[dddrr,bend left=50,"j"]\\
      B\ar[r,"g"']\ar[ddrrr,bend right=50,"g"']&B'\ar[ddrr,"\id_{B'}"]\\
      &&\phantom{A'}\\
      &&&B'
    \end{tikzcd}
  \]
  This shows that $j\circ j'=\id_{B'}$. Similarly, we have $j'\circ j=\id_{A'}$
  and $j$ is thus an isomorphism. Consider a countable composition of
  isomorphisms $f_i:A_i\to A_{i+1}$ as in~\eqref{eq:ccomp}. There is a cocone
  on~$A_0$ consisting of the morphisms
  \[
    f_0^{-1}\circ f_1^{-1}\circ\ldots\circ f_{i-1}^{-1}:A_i\to A_0
  \]
  which is easily seen to be universal and the composite is thus (isomorphic to)
  $\id_{A_0}$.
  Consider a retract $j$ of an isomorphism $i$ as in \eqref{eq:retract}. We
  claim that the morphism $j'=s\circ i^{-1}\circ r'$ is the inverse
  of~$j$. Namely, one has
  \begin{align*}
    j'\circ j
    &=s\circ i^{-1}\circ r'\circ j
    &
    j\circ j'
    &= j\circ s\circ i^{-1}\circ r'
    \\
    &=s\circ i^{-1}\circ i\circ r
    &
    &=s'\circ i\circ i^{-1}\circ r'
    \\
    &=s\circ r
    &
    &=s'\circ r'
    \\
    &=\id_{A'}
    &
    &=\id_{B'}
    \qedhere
  \end{align*}
\end{proof}

\subsection{Generating cofibrations}
\label{sec:pres-gen-cof}
Consider the presentation with one generator and no relation:
\[
  \presG
  =
  \pres{a}{}
  \pbox.
\]
Given $m,n\in\N$, we introduce notations for the following presentations,
respectively with $n$~generators and no relation, and with one relation between
words of respective lengths $m$ and $n$:
\begin{align*}
  \presG^n
  &=
  \pres{a_1,\ldots,a_n}{}
  &
  \presR mn
  &=
  \pres{a_1,\ldots,a_{m+n}}{a_1\ldots a_m\rto a_{m+1}\ldots a_{m+n}}
  \pbox.
\end{align*}
%
We write $\clI$ for the class of morphisms, called \emph{generating
  cofibrations}, consisting of the obvious inclusions of presentations
\begin{align*}
  g:\iobj&\into\presG
  &
  r^{m,n}:\presG^{m+n}&\into\presR mn
\end{align*}
for some $m,n\in\N$.

\subsection{Cofibrations}
\label{sec:pres-cof}
We write $\clC=\lrlp\clI$ for the class of morphisms whose elements are
called \emph{cofibrations}.
Note that, given a presentation~$\P$, the pushouts
\[
  \begin{tikzcd}
    \iobj\ar[d]\ar[r,"g"]\ar[phantom,dr,"\ulcorner",very near end]&\presG\ar[d,dotted]\\
    \P\ar[r,dotted]&\Q
  \end{tikzcd}
  \qquad\qquad\qquad
  \begin{tikzcd}
    \presG^{m+n}\ar[d,"f"']\ar[r,"r^{m,n}"]\ar[phantom,dr,"\ulcorner",very near end]&\presR mn\ar[d,dotted]\\
    \P\ar[r,dotted]&\Q
  \end{tikzcd}
\]
are respectively the presentation obtained from~$\P$ by adding a generator and the
presentation obtained from~$\P$ by adding a relation (between the two words of
$\freemon\P_1$ specified by~$f$).

\begin{lemma}
  Every presentation~$\P$ is cofibrant, in the sense that the initial morphism
  $\iobj\into\P$ is a cofibration.
\end{lemma}
\begin{proof}
  By \propr{small-obj}, it is enough to show that the initial morphism
  $\iobj\into\P$ can be obtained as a composite of pushouts of generating
  cofibrations, which amounts to show that every presentation can be obtained
  from the empty one by adding generators and relations, which we will do in
  this order (generators first, and then relations). Given a relation
  $u\rto v\in\P_2$, we have a canonical inclusion
  \[
    \begin{tikzcd}
      \presG^{\wlen u+\wlen v}\ar[r,"r^{\wlen u+\wlen v}"]&\presR{\wlen u}{\wlen v}
    \end{tikzcd}
  \]
  and a canonical inclusion
  \[
    \begin{tikzcd}
    \presG^{\wlen u+\wlen v}\ar[r]&\coprod_{a\in\P_1}\presG\pbox.      
    \end{tikzcd}
  \]
  By summing those morphisms over relations $(u,v)\in\P_2$, and post-composing
  with the codiagonal
  \[
    \begin{tikzcd}
      \coprod_{(u,v)\in\P_2}\coprod_{a\in\P_1}\presG\ar[r]&\coprod_{a\in\P_1}\presG\pbox,
    \end{tikzcd}
  \]
  we obtain a diagram
  \[
    \begin{tikzcd}
      \coprod_{(u,v)\in\P_2}\presG^{\wlen u+\wlen v}\ar[d]\ar[r]&\coprod_{(u,v)\in\P_2}\presR{\wlen u}{\wlen v}&\\
      \coprod_{a\in\P_1}\presG
    \end{tikzcd}
  \]
  whose pushout is precisely~$\P$. Finally, we consider the composite of morphisms
  \[
    \begin{tikzcd}
      \iobj\ar[r]&\coprod_{a\in\P_1}\presG\ar[r]&\P
    \end{tikzcd}
  \]
  where the second morphism is constructed in the cocone of the pushout. Again,
  this composite expresses the fact that any presentation can be constructed
  from the empty one by first adding all the generators, and then adding all the
  relations.
\end{proof}

\noindent
The construction given in the above proof easily generalizes to show:

\begin{lemma}
  \label{lem:mon-cof}
  Any monomorphism $f:\P\to\Q$ is a cofibration (and, in fact, an
  $\clI$-cellular extension).
\end{lemma}

\noindent
Conversely, one has:

\begin{lemma}
  \label{lem:cof-mon}
  Cofibrations are monomorphisms.
\end{lemma}
\begin{proof}
  The generating cofibrations are monomorphisms. Moreover, monomorphisms are
  closed under coproducts, under pushouts and countable compositions by
  \lemr{mono-stab}. By \propr{small-obj}, cofibrations are thus retracts of
  monomorphisms. We conclude using the fact that monomorphisms are closed under
  retracts.
  Namely, suppose given a retract $j$ of a monomorphism~$i$, as
  in~\eqref{eq:retract}, and two morphisms $h_1,h_2$ such that
  $j\circ h_1=j\circ h_2$, we have
  \begin{align*}
    j\circ h_1&=j\circ h_2\\
    s'\circ j\circ h_1&=s'\circ j\circ h_2\\
    i\circ s\circ h_1&=i\circ s\circ h_2\\
    s\circ h_1&=s\circ h_2\\
    r\circ s\circ h_1&=r\circ s\circ h_2\\
    h_1&=h_2
  \end{align*}
  and we conclude.
\end{proof}

\begin{corollary}
  The class~$\C$ of cofibrations is the class of monomorphisms in~$\rPres$.
\end{corollary}

\subsection{Trivial fibrations}
The morphisms in the class $\rlp\clI$ are called \emph{trivial fibrations}. From
the lifting property with respect to the generators we immediately deduce,

\begin{lemma}
  \label{lem:tfib}
  The morphisms $f:\P\to\Q$ in $\rlp\clI$ are those
  \begin{itemize}
  \item whose underlying function $f:\P_1\to\Q_1$ is surjective, and
  \item such that for every $u,v\in\freemon\P_1$,
    $\freemon f(u)\rto \freemon f(v)\in\Q_2$ implies $u\rto v\in\P_2$.
  \end{itemize}
\end{lemma}


\begin{lemma}
  \label{lem:pres-tfib-w}
  Trivial fibrations are weak equivalences: $\rlp\clI\subseteq\clW$.
\end{lemma}
\begin{proof}
  Since $f:\P_1\to\Q_1$ is surjective, we have that $\pmon f:\pmon\P\to\pmon\Q$
  is also surjective. We have to show that it is also injective in order to
  conclude. Suppose given $u,v\in\freemon\P_1$ such that
  $\freemon f(u)\psim\Q\freemon f(v)$: we have a sequence
  \[
    \freemon f(u)=w_0\rtot w_1\rtot \ldots\rtot w_n=\freemon f(v)
  \]
  where the arrows ``$\rtot$'' mean that, for $0\leq i<n$, there is a
  decomposition of $w_i$ and $w_{i+1}$ as
  \[
    w_i=t_iu_iv_i
    \quad\text{and}\quad
    w_{i+1}=t_i'u_i'v_i'
    \qquad\text{with}\qquad
    u_i\rto u_i'\in\Q_2
    \quad\text{or}\quad
    u_i\rot u_i'\in\Q_2
    \pbox.
  \]
  Moreover, since $\Q$ is reflexive, we can always suppose that this sequence is
  non-empty, \ie $n>0$: we can replace the empty sequence by the reflexivity
  relation $\freemon f(u)\rto\freemon f(u)$.
  By surjectivity, for $0\leq i\leq n$, there are words $t_i^\P$, $u_i^\P$,
  $v_i^\P$, $t_i'^\P$, $u_i'^\P$, $v_i'^\P$ in $\freemon\P_1$ whose image
  under~$f$ is respectively $t_i$, $u_i$, $v_i$, $t_i'$, $u_i'$, $v_i'$, and we
  may moreover assume $t_0^\P u_0^\P v_0^\P=u$ and
  $t_{n-1}'^\P u_{n-1}'^\P v_{n-1}'^\P=v$. Finally, since $f$ is a trivial
  fibration, we have $u_i\rto u_i'$ or $u_i\rot u_i'$ and we conclude
  that~$u\psim\P v$.
\end{proof}


\noindent
From the results of \secr{pres-epi}, one has:

\begin{lemma}
  Every trivial fibration is an epimorphism.
\end{lemma}

\subsection{Trivial cofibrations}
The class of \emph{trivial cofibrations} is $\clC\cap\clW$ and consists of
monomorphisms $f:\P\to\Q$ such that the induced morphism of monoids
$\pmon f:\pmon\P\to\pmon\Q$ is an isomorphism.

\begin{lemma}
  A morphism $f:\P\to\Q$ is a trivial cofibration when
  \begin{itemize}
  \item $f$ is a monomorphism,
  \item for every $a\in\Q_1$, there exists $u\in\freemon\P_1$ such that
    $f(u)\psim\Q a$,
  \item for $u,v\in\freemon\P_1$ such that $f(u)\psim\Q f(v)$, we have
    $u\psim\P v$.
  \end{itemize}
\end{lemma}
\begin{proof}
  Suppose that $f$ is a trivial cofibration.
  Since $f$ is a cofibration, it is a monomorphism by \lemr{cof-mon}.
  Given $a\in\Q_1$, since $\pmon f$ is surjective there is $u\in\freemon\P_1$
  such that $f(\pmon u)=\pmon a$, and we have $f(u)\psim\Q a$.
  %
  Given $u,v\in\freemon\P_1$ such that $f(u)\psim\Q f(v)$, we have
  $\pmon f(\pmon u)=\pmon f(\pmon v)$, thus $\pmon u=\pmon v$ in $\pmon\P$ since
  $\pmon f$ injective, and finally~$u\psim\P v$.

  Conversely, suppose given a monomorphism $f:\P\to\Q$.
  Given $a\in\Q_1$, by hypothesis, there exists $u_a\in\freemon\P_1$ such that
  $\pmon f(\pmon u_a)=\pmon a$. Therefore, given $\pmon v\in\pmon\Q$, for some
  $v=a_1\ldots a_n\in\freemon\Q_1$, we have
  \[
    \pmon f(\pmon u_{a_1}\ldots\pmon u_{a_n})=\pmon f(\pmon u_{a_1})\ldots\pmon f(\pmon u_{a_n})=\pmon a_1\ldots\pmon a_n=\pmon v
  \]
  and $\pmon f$ is thus surjective.
  %
  Suppose given $u,v\in\freemon\P_1$ such that
  $\pmon f(\pmon u)=\pmon f(\pmon v)$: we have $f(u)\psim\Q f(v)$, thus
  $u\psim\P v$ and finally $\pmon u=\pmon v$.
\end{proof}

\begin{lemma}
  \label{lem:cofib-clos}
  The class of trivial cofibrations satisfies
  $\lrlp{(\clC\cap\clW)}=\clC\cap\clW$.
\end{lemma}
\begin{proof}
  By \lemr{ortho}, we have
  $\clC\cap\clW\subseteq\lrlp{(\clC\cap\clW)}$. Conversely, by
  \cref{prop:small-obj}, every element of $\lrlp{(\clC\cap\clW)}$ is a retract
  of a $(\clC\cap\clW)$-cellular extension, and thus belongs to $\clC\cap\clW$,
  because this last class is closed under sums, pushouts, countable compositions
  and retracts by \cref{lem:llp-stab,lem:pres-w-stab}.
\end{proof}

\subsection{Fibrations}
The class $\clF$ of \emph{fibrations} is determined by the two other classes:
should there be a model structure, it is necessarily
$\clF=\rlp{(\clC\cap\clW)}$. An explicit description of fibrant objects is given
by \cref{lem:pfib-fib-obj,lem:pfib}.

\subsection{Toward a model structure}

We now have almost all the ingredients required to construct a model structure
on the category~$\rPres$ of reflexive presentations with~$\clW$ as weak
equivalences, $\clC=\lrlp\clI$ as cofibrations and $\clF=\rlp{(\clC\cap\clW)}$
as fibrations. Namely, assuming that $\clJ=\clC\cap\clW$ is a set, we can
apply~\propr{cof-gen}, whose hypothesis can be shows as follows.
Closure under retracts and the 2-out-of-3 property for $\clW$ were shown in
\lemr{pres-w-23}.
We first show $\rlp\clI\subseteq\rlp\clJ\cap\clW$. We have
$\clJ=\clC\cap\clW\subseteq\clC$ thus, by \cref{lem:ortho},
$\rlp\clI=\rlp{(\llp{(\rlp\clI)})}=\rlp\clC\subseteq\rlp\clJ$; and we have
$\rlp\clI\subseteq\clW$ by \cref{lem:pres-tfib-w}.
Finally, the remaining inclusion is shown by \lemr{cofib-clos}, since we have
\[
  \lrlp\clJ=\lrlp{(\clC\cap\clW)}=\clC\cap\clW=\lrlp\clI\cap\clW
\]
which concludes the proof.

However, it is not clear at all that the class $\clJ$ should be a set. This is
why our actual construction of a model category uses Smith's theorem, which
provides sufficient conditions in order to ensure that there exists a set~$J$
which can be used as a replacement for~$\clJ$, in the sense that we have
$\lrlp J=\clJ$.

\subsection{Recognizing weak equivalences}
In preparation for the use of Smith's theorem, we show that morphisms of
presentations which are weak equivalences can be characterized by factorization
properties as follows. First note that the words in a given presentation can be
represented in the following way.

\begin{lemma}
  \label{lem:repr-word}
  Given a natural number $n$ and a presentation~$\P$, there is a bijection
  between morphisms $\presG^n\to\P$ and words in $\freemon{\P_1}$ of length~$n$.
\end{lemma}
\begin{proof}
  To a word $u=b_1\ldots b_n$ of length $n$, we associate the morphism
  $f:\presG^n\to\P$ such that $f(a_i)=b_i$ for $1\leq i\leq n$. Conversely, a
  morphism $f:\presG^n\to\P$, we associate the word $u=f(a_1)\ldots f(a_n)$. The
  two operations are easily seen to be inverse of each other.
\end{proof}

\begin{lemma}
  \label{lem:repr-word-im}
  Suppose given a morphism $w:\P\to\Q$ and $u\in\freemon\P_1$. Writing
  $f:\presG^{\wlen u}\to\P$ for the morphism associated to $u$ by
  \cref{lem:repr-word}, the morphism associated to $w\circ f$ by
  \cref{lem:repr-word} is $\freemon w(u)$.
\end{lemma}
\begin{proof}
  Direct by inspection of the bijection constructed in the proof of
  \cref{lem:repr-word}.
\end{proof}

\noindent
Similarly, we can represent pairs of words as follows, where we write
$\presG^{m,n}$ instead of $\presG^{m+n}$:

\begin{lemma}
  \label{lem:repr-words}
  Given natural numbers $m,n$ and a presentation~$\P$, there is a bijection
  between morphisms $\presG^{m,n}\to\P$ and pairs of words
  $u,v\in\freemon{\P_1}$ with $\wlen{u}=m$ and $\wlen{v}=n$.
\end{lemma}
\begin{proof}
  A word $u=b_1\ldots b_mb_{m+1}\ldots b_{m+n}$ of length $m+n$ in
  $\freemon\P_1$ can be seen as the pair of words $b_1\ldots b_m$ and
  $b_{m+1}\ldots b_{m+n}$ in $\freemon\P_1$ of respective lengths $m$ and
  $n$. Thus the result by \cref{lem:repr-word}.
\end{proof}

\newcommand{\presE}{\presentation{E}}

\noindent
We now construct a family of presentations to represent pairs of equivalent
words. Suppose fixed natural numbers~$m$ and~$n$. Given a natural number~$o$,
consider the set
\[
  G=\set{a_1,\ldots,a_m,b_1,\ldots,b_n,c_1,\ldots,c_o}
\]
Suppose moreover given a non-zero natural number $k$ and words
$u_i,v_i,v_i',w_i\in\freemon G$ with $1\leq i\leq k$, such that
\begin{align*}
  u_1v_1w_1&=a_1\ldots a_n
  &
  u_iv_i'w_i&=u_{i+1}v_{i+1}w_{i+1}
  &
  u_kv_kw_k&=b_1\ldots b_n
\end{align*}
for every index $i$ with $1\leq i<n$. We write $\presE^{m,n}$ for a presentation
of the form
\[
  \presE^{m,n}=\pres{G}{v_1\rtot v_1',\ldots,v_k\rtot v_k'}
\]
where $v_i\rtot v_i'$ is either $v_i\rto v_i'$ or $v_i'\rto v_i$, called an
\emph{equivalence presentation}. Note that there is a canonical inclusion
$e:\presG^{m,n}\to\presE^{m,n}$ such that $e(a_i)=a_i$ for $1\leq i\leq m$ and
$e(a_{m+i})=b_i$ for~$1\leq i\leq n$. Also note that $\presR mn$ is a particular
case of $\presE^{m,n}$ (where $k=1$).
We could further reduce the number of equivalence presentations that we consider
by imposing conditions such as the fact that there is no repeated generator
within each $v_i$ or $v_i'$, that $v_i$ and $v_i'$ do not share any common
generator for every index $i$, that each generator $c_j$ occurs within $v_i$ or
$v_i'$ for some index~$i$, and so on, but this will play no significant role in
the following.

\begin{lemma}
  \label{lem:repr-equiv}
  Suppose given a presentation~$\P$ and a pair of words $u,v\in\freemon\P_1$
  with $\wlen{u}=m$ and $\wlen{v}=n$, corresponding to a morphism
  $f:\presG^{m,n}\to\P$ via \cref{lem:repr-words}. We have $u\psim\P v$ if and
  only if there exists a morphism $g:\presE^{m,n}\to\P$ making the following
  diagram commute, for some equivalence presentation~$\presE^{m,n}$,
  \[
    \begin{tikzcd}
      \presG^{m,n}\ar[d]\ar[dr,"f"]\\
      \presE^{m,n}\ar[dotted,r,"g"']&\P
    \end{tikzcd}
  \]
  where the vertical arrow is the canonical inclusion.
\end{lemma}

\begin{proposition}
  \label{prop:surj-fact}
  Suppose given a morphism of presentations $w:\P\to\Q$. The induced
  morphism $\pmon w:\pmon\P\to\pmon\Q$ is surjective if and only if every square
  as on the left factors as on the right
  \[
    \begin{tikzcd}
      \iobj\ar[d]\ar[r]&\P\ar[d,"w"]\\
      \presG\ar[r,"g"']&\Q
    \end{tikzcd}
    \qquad\qquad\qquad\qquad
    \begin{tikzcd}
      \iobj\ar[d]\ar[rr,bend left]\ar[r]&\presG^n\ar[d,"h"description]\ar[r,"f''"']&\P\ar[d,"w"]\\
      \presG\ar[rr,bend right,"g"']\ar[r,"g'"]&\presE^{1,n}\ar[r,"g''"]&\Q
    \end{tikzcd}
  \]
  where $g':\presG\to\presE^{1,n}$ sends $a$ to $a_1$ and
  $h:\presG^n\to\presE^{1,n}$ sends $a_i$ to $b_i$.
\end{proposition}
\begin{proof}
  The presentation $\iobj$ being initial, the square on the left is uniquely
  determined by $g$ which, by \cref{lem:repr-word}, corresponds to a
  generator~$a$ in $\Q_1$. The diagram on the right corresponds to the existence
  of a word~$v$ of length~$n$ in $\freemon\P_1$ (given by $f''$ via
  \cref{lem:repr-word}) such that $a\psim\Q\freemon{w}(v)$ (this is given
  by~$g''$ via \cref{lem:repr-word-im,lem:repr-equiv}). The above factorization
  property thus amounts to requiring that every generator $a$ in $\Q_1$ is
  equivalent to some word of the form $\freemon w(v)$ for some
  $v\in\freemon\P_1$, and thus that every word $u\in\freemon\Q_1$ is equivalent
  to some word of the form $\freemon w(v)$ for some $v\in\freemon\P_1$, \ie that
  $\pmon w$ is surjective.
\end{proof}

\begin{proposition}
  \label{prop:inj-fact}
  Suppose given a morphism of presentations $w:\P\to\Q$. The induced morphism
  $\pmon w:\pmon\P\to\pmon\Q$ is injective if and only if every square as on the
  left (where the vertical morphism $i:\presG^{m,n}\to\presE^{m,n}$ is the
  canonical inclusion into some equivalence presentation) factors as on the
  right, where $\presE_1^{m,n}$ is an equivalence presentation (the index
  stresses the fact that it might be different from $\presE^{m,n}$) and the
  square on the left is a pushout:
  \[
    \begin{tikzcd}
      \presG^{m,n}\ar[d,"i"']\ar[r,"f"]&\P\ar[d,"w"]\\
      \presE^{m,n}\ar[r,"g"']&\Q
    \end{tikzcd}
    \qquad\qquad\qquad\qquad
    \begin{tikzcd}
      \presG^{m,n}\ar[d,"i"']\ar[rr,bend left,"f"]\ar[r,"f'"']\ar[dr,phantom,very near end,"\ulcorner"]&\presE_1^{m,n}\ar[d]\ar[r,"f''"']&\P\ar[d,"w"]\\
      \presE^{m,n}\ar[rr,bend right,"g"']\ar[r,"g'"]&\presE\ar[r,"g''"]&\Q
    \end{tikzcd}
  \]
\end{proposition}
\begin{proof}
  The square on the left amounts to giving two words $u$ and $v$ in
  $\freemon\P_1$ of respective lengths $m$ and~$n$ (by $f$ via
  \cref{lem:repr-words}) such that $\freemon w(u)\psim\Q\freemon w(v)$ (by $g$
  via \cref{lem:repr-word-im,lem:repr-equiv}). The diagram on the right
  corresponds to supposing that we have $u\psim\P v$ (by $f''$ via
  \cref{lem:repr-equiv}) (note that $g''$ does not bring any information by
  the universal property of the pushout). The above factorization property thus
  amounts to requiring that for every words $u,v\in\freemon\P_1$ such that
  $\freemon w(u)\psim\Q\freemon w(v)$, we have $u\psim\P v$, \ie that
  $\pmon w:\pmon\P\to\pmon\Q$ is injective.
\end{proof}

A presentation~$\P$ is \emph{countable} when the set $\P_1$ is countable. We
write $\clW^{\leq\omega}$ for the class of weak equivalences $w:\P\to\Q$ such
that both $\P$ and $\Q$ are countable. Up to isomorphism, we can always suppose
that we have $\P_1\subseteq\N$ and $\Q_1\subseteq\N$, so that
$\clW^{\leq\omega}$ is essentially a set (as opposed to a class).

\begin{proposition}
  \label{prop:solution-set}
  Every commutative square as on the left, where $\R$ and $\S$ are finite and
  $w$ is a weak equivalence, factors as a square as on the right with
  $w^\omega\in\clW^{\leq\omega}$:
  \[
    \begin{tikzcd}
      \R\ar[d,"i"']\ar[r,"f"]&\P\ar[d,"w"]\\
      \S\ar[r,"g"']&\Q
    \end{tikzcd}
    \qquad\qquad\qquad\qquad
    \begin{tikzcd}
      \R\ar[d,"i"']\ar[rr,bend left,"f"]\ar[r,"f'"']&\P^\omega\ar[d,"w^\omega"description]\ar[r,"f''"']&\P\ar[d,"w"]\\
      \S\ar[rr,bend right,"g"']\ar[r,"g'"]&\Q^\omega\ar[r,"g''"]&\Q
    \end{tikzcd}
  \]
\end{proposition}
\begin{proof}
  We are going to construct a sequence $w^k:\P^k\to\Q^k$ of morphisms, indexed
  by $k\in\N$, such that both $\P^k$ and $\Q^k$ are finite, the left square above
  factors through each $w^k$, \ie
  \[
    \begin{tikzcd}
      \R\ar[d,"i"']\ar[rr,bend left,"f"]\ar[r,"f'_k"']&\P^k\ar[d,"w^k"description]\ar[r,"f''_k"']&\P\ar[d,"w"]\\
      \S\ar[rr,bend right,"g"']\ar[r,"g'_k"]&\Q^k\ar[r,"g''_k"]&\Q
    \end{tikzcd}
  \]
  we have $\P^k\subseteq\P^{k+1}$ and $\Q^k\subseteq\Q^{k+1}$, and the morphisms
  $w^k$ as well as the factorizations respect those inclusions. Since $\P^k$ and
  $\Q^k$ are finite, it will be the case that we can moreover suppose that
  $\P_1^k$ and $\Q_1^k$ are initial segments of~$\N$, so that it makes sense to
  consider the smallest generator or relation satisfying a property. We first
  define $w^0$ to be a morphism isomorphic to $i$, such that both $\P_1^0$ and
  $\Q_1^0$ are initial segments of~$\N$. Suppose $w^k$ defined, we alternate
  between the two operations below in order to construct $w^{k+1}$.
  \begin{enumerate}[(A)]
  \item\label{surj-step} Consider a generator $a\in\Q^k$ which has no antecedent
    under~$\pmon w^k:\pmon\P^k\to\pmon\Q^k$. By \cref{lem:repr-word}, it
    corresponds to a morphism~$\presG\to\Q^k$. By \cref{prop:surj-fact}, since
    $w$ is surjective, the outer square of the diagram on the left factors as on
    the right, for some equivalence presentation~$\presE^{1,n}$:
    \[
      \begin{tikzcd}
        \iobj\ar[d]\ar[r]&\P^k\ar[d,"w^k"description]\ar[r,"f''_k"]\ar[phantom,dr,"\text{(a)}"]&\P\ar[d,"w"]\\
        \presG\ar[r]&\Q^k\ar[r,"g''_k"']&\Q
      \end{tikzcd}
      \qquad\qquad\qquad\qquad
      \begin{tikzcd}
        \iobj\ar[d]\ar[r]&\presG^n\ar[d]\ar[r]\ar[phantom,dr,"\text{(b)}"]&\P\ar[d,"w"]\\
        \presG\ar[r]&\presE^{1,n}\ar[r]&\Q
      \end{tikzcd}
    \]
    We define the pushouts $\P^{k+1}=\P^k\sqcup\presG^n$ and
    $\Q^{k+1}=\Q^k\sqcup_\presG\presE^{1,n}$, and the morphism
    $w^{k+1}:\P^{k+1}\to\Q^{k+1}$ is defined by the expected universal property
    of $\P^{k+1}$, see the diagram on the left below:
    \[
      \begin{tikzcd}[sep=1ex]
        \iobj\ar[dd]\ar[dr]\ar[rr]\ar[phantom,drrr,"\ulcorner",very near end]&&\presG^n\ar[dd]\ar[dr,dotted]\\
        &\P^k\ar[dd,"w^k" near start]\ar[rr,dotted]&&\P^{k+1}\ar[dd,"w^{k+1}",dotted]\\
        \presG\ar[dr]\ar[rr]\ar[phantom,drrr,"\ulcorner",very near end]&&\presE^{1,n}\ar[dr,dotted]\\
        &\Q^k\ar[rr,dotted]&&\Q^{k+1}
      \end{tikzcd}
      \qquad\qquad\qquad
      \begin{tikzcd}
        \P^k\ar[d,"w^k"']\ar[r]&\P^{k+1}\ar[d,"w^{k+1}"description]\ar[r,dotted,"f''_{k+1}"]\ar[phantom,dr,"\text{(c)}"]&\P\ar[d,"w"]\\
        \Q^k\ar[r]&\Q^{k+1}\ar[r,dotted,"g''_{k+1}"']&\Q
      \end{tikzcd}
    \]
    We can construct the pushouts so that $\P^k\subseteq\P^{k+1}$,
    $\Q^k\subseteq\Q^{k+1}$ and both $\P_1^{k+1}$ and~$\Q_1^{k+1}$ are initial
    segments of~$\N$.
    Finally, the horizontal morphisms of the diagram (c) are defined from (a)
    and (b) by universal property of the pushouts, and the outer square on the
    right above is a factorization of (a).
  \item\label{inj-step} Consider a pair of words $u,v\in\freemon{(\P^k)_1}$
    such that there is an equivalence $\freemon{(w^k)}(u)\psim\Q\freemon{(w^k)}(v)$ and
    $\pmon u\neq\pmon v$ in $\pmon\P^k$. By \cref{lem:repr-words}, the pair of
    words $(u,v)$ in $\P^k$ corresponds to a morphism $\presG^{m,n}\to\P^k$ and,
    by \cref{lem:repr-equiv}, the equivalence
    $\freemon{(w^k)}(u)\psim\Q\freemon{(w^k)}(v)$ to a morphism $\presE^{m,n}\to\Q$, where
    $m$ and $n$ are the respective lengths of~$u$ and~$v$. By
    \cref{prop:inj-fact}, the outer square on the left (where the morphism
    $\presG^{m,n}\to\presE^{m,n}$ is the canonical inclusion) factors as on the
    right,
    \[
      \begin{tikzcd}
        \presG^{m,n}\ar[d]\ar[r]&\P^k\ar[d,"w^k"description]\ar[r,"f''_k"]\ar[phantom,dr,"\text{(d)}"]&\P\ar[d,"w"]\\
        \presE^{m,n}\ar[r]&\Q^k\ar[r,"g''_k"']&\Q
      \end{tikzcd}
      \qquad\qquad\qquad\qquad
      \begin{tikzcd}
        \presG^{m,n}\ar[d]\ar[r]\ar[phantom,dr,"\ulcorner",very near end]&\presE_1^{m,n}\ar[d]\ar[r]\ar[phantom,dr,"\text{(e)}"]&\P\ar[d,"w"]\\
        \presE^{m,n}\ar[r]&\presE\ar[r]&\Q
      \end{tikzcd}
    \]
    We define the pushouts $\P^{k+1}=\P^k\sqcup_{\presG^{m,n}}\presE_1^{m,n}$ and
    $\Q^{k+1}=\Q^k\sqcup_{\presE^{m,n}}\presE$, and the morphism
    $w^{k+1}:\P^{k+1}\to\Q^{k+1}$ is defined by the expected universal property
    of $\P^{k+1}$, see the diagram on the left below:
    \[
      \begin{tikzcd}[sep=1ex]
        \presG^{m,n}\ar[dd]\ar[dr]\ar[rr]\ar[phantom,drrr,"\ulcorner",very near end]&&\presE_1^{m,n}\ar[dd]\ar[dr,dotted]\\
        &\P^k\ar[dd,"w^k" near start]\ar[rr,dotted]&&\P^{k+1}\ar[dd,"w^{k+1}",dotted]\\
        \presE^{m,n}\ar[dr]\ar[rr]\ar[phantom,drrr,"\ulcorner",very near end]&&\presE\ar[dr,dotted]\\
        &\Q^k\ar[rr,dotted]&&\Q^{k+1}
      \end{tikzcd}
      \qquad\qquad\qquad
      \begin{tikzcd}
        \P^k\ar[d,"w^k"']\ar[r]&\P^{k+1}\ar[d,"w^{k+1}"description]\ar[r,dotted,"f''_{k+1}"]\ar[phantom,dr,"\text{(f)}"]&\P\ar[d,"w"]\\
        \Q^k\ar[r]&\Q^{k+1}\ar[r,dotted,"g''_{k+1}"']&\Q
      \end{tikzcd}
    \]
    We can construct the pushouts so that $\P^k\subseteq\P^{k+1}$,
    $\Q^k\subseteq\Q^{k+1}$ and both $\P_1^{k+1}$ and~$\Q_1^{k+1}$ are initial
    segments of~$\N$.
    Finally, the horizontal morphisms of the diagram (f) are defined from (d)
    and (e) by universal property of the pushouts, and the outer square on the
    right above is a factorization of (d).
  \end{enumerate}
  By alternatively using \ref{surj-step} and \ref{inj-step}, we construct a
  sequence of $w^k:\P^k\to\Q^k$ of morphisms of presentations, such that every
  generator of $\Q^k_1$ is eventually handled by \ref{surj-step} and every
  relation in $\Q^k$ between words in the image is eventually handled by
  \ref{inj-step}.
  More explicitly, this can be performed as follows. We say that a generator
  $a\in\P^k_1$ has \emph{appeared} at step $i$ when $(u,v)\in\P^i_1$ and
  $a\not\in\P^{i-1}_1$; more generally, we say that a word
  $u\in\freemon{(\P^k)}$ has appeared at step $i$ if it contains a generators
  which has appeared at step~$i$ and all the generators it contains have appeared
  at step~$j\geq i$; we say that a pair $(u,v)$ of words have appeared at
  step~$i$ if $u$ has appeared at step $i$ and $v$ has appeared at step
  $j\geq i$, or the converse. We then iteratively perform the following steps in
  order to define the terms of the sequence $(w^k)_{k\in\N}$: supposing that
  $w^k$ is defined, we defined $w^{k+1}$ to $w^{2k+2}$ as follows.
  \begin{itemize}
  \item We construct $w^{k+1}$ by using~\ref{surj-step} on the smallest generator
    $a$ which is not in the image of~$\pmon{w^k}$ (we take $w^{k+1}=w^k$ if no
    such generator exists).
  \item We construct $w^{k+1+i+1}$, for $0\leq i\leq k$, where $w^{k+1+i+1}$ is
    defined from $w^{k+1+i}$ by using~\ref{inj-step} on the smallest pair of
    words $u,v\in\freemon{(P^k)}_1$ which has appeared at step~$i$, such that
    there is an equivalence $\freemon{(w^k)}\psim\Q\freemon{(w^k)}$ and
    $\pmon u\neq\pmon v\in\P^k$ (we take $w^{k+1+i+1}=w^{k+1+i}$ if no such pair
    exists).
  \end{itemize}
  Finally, we define the colimits $\P^\omega=\bigcup_{k\in\N}\P^k$ and
  $\Q^\omega=\bigcup_{k\in\N}\Q^k$, which are countable presentations as unions
  of finite ones, and $w^\omega:\P^\omega\to\Q^\omega$ as the morphism induced
  by the cocone consisting of the morphisms $w^k:\P^k\to\Q^k\subseteq\Q^\omega$.
  Given a square
  \[
    \begin{tikzcd}
      \iobj\ar[d]\ar[r]&\P^\omega\ar[d,"w^\omega"]\\
      \presG\ar[r,"g"']&\Q^\omega
    \end{tikzcd}
  \]
  the generator corresponding to the morphism $g:\presG\to\Q$ was handled by
  \ref{surj-step} at some step $k$ (or had a lifting from the beginning, this
  case being simple). There is therefore a factorization of the square on the
  left as on the middle
  \[
    \begin{tikzcd}
      \iobj\ar[d]\ar[r]&\P^{k+1}\ar[d,"w^{k+1}"]\\
      \presG\ar[r,"g"']&\Q^{k+1}
    \end{tikzcd}
    \qquad\qquad\qquad\qquad
    \begin{tikzcd}
      \iobj\ar[d]\ar[rr,bend left]\ar[r]&\presG^n\ar[d,"h"description]\ar[r,"f''"']&\P^{k+1}\ar[d,"w^{k+1}"]\\
      \presG\ar[rr,bend right,"g"']\ar[r,"g'"]&\presE^{1,n}\ar[r,"g''"]&\Q^{k+1}
    \end{tikzcd}
  \]
  for some equivalence presentation $\presE^{1,n}$, from which we deduce the factorization
  \[
    \begin{tikzcd}
      \iobj\ar[d]\ar[rr,bend left]\ar[r]&\presG^n\ar[d,"h"description]\ar[r,"f''"']&\P^{k+1}\ar[d,"w^{k+1}"]\ar[r]&\P^\omega\ar[d,"w^\omega"]\\
      \presG\ar[rr,bend right,"g"']\ar[r,"g'"]&\presE^{1,n}\ar[r,"g''"]&\Q^{k+1}\ar[r]&\Q^\omega
    \end{tikzcd}
  \]
  of the original square by post-composing with the canonical inclusions into
  the colimit. By \cref{prop:surj-fact}, the morphism $\pmon w^\omega$ is thus
  surjective.
  Similarly, by considering steps \ref{inj-step} and \cref{prop:inj-fact}, one
  can show that $\pmon w^\omega$ is surjective. The morphism $w^\omega$ is thus
  a weak equivalence as desired.
\end{proof}

\subsection{A model structure}
Given a class $\clW$ of morphisms and a morphism $i$, we say that $\clW$
satisfies the \emph{solution set condition} at $i$ if there is a set
$\clW_i\subseteq\clW$ of morphisms such that any commutative square as on the
left, with $w\in\clW$, factors as a square as on the right, for some
$w'\in\clW_i$:
\begin{equation}
  \label{eq:solution-set}
  \begin{tikzcd}
    A\ar[d,"i"']\ar[r,"f"]&X\ar[d,"w"]\\
    B\ar[r,"g"']&Y
  \end{tikzcd}
  \qquad\qquad\qquad\qquad
  \begin{tikzcd}
    A\ar[d,"i"']\ar[rr,bend left,"f"]\ar[r,"f'"']&X'\ar[d,"w'"description]\ar[r,"f''"']&X\ar[d,"w"]\\
    B\ar[rr,bend right,"g"']\ar[r,"g'"]&Y'\ar[r,"g''"]&Y
  \end{tikzcd}
\end{equation}
By extension, given a set $\clI$ of morphisms, we say that $\clW$ satisfies the
solution set condition at~$\clI$, if it satisfies the solution set condition at
any $i\in\clI$. The following theorem is due to Smith,
see~\cite[Theorem~1.7]{beke2000sheafifiable}:

\begin{theorem}
  \label{thm:smith}
  In a locally finitely presentable category, suppose given a subcategory~$\clW$
  and a set~$\clI$ of morphisms such that
  \begin{enumerate}
  \item $\clW$ is closed under retracts and has the 2-out-of-3 property,
  \item $\rlp\clI\subseteq\clW$,
  \item $\lrlp\clI\cap\clW$ is closed under pushouts and countable compositions,
  \item $\clW$ satisfies the solution set condition at~$\clI$.
  \end{enumerate}
  Then there is a cofibrantly generated model structure with $\lrlp\clI$ as
  cofibrations, $\clW$ as weak equivalences and $\rlp{(\lrlp\clI\cap\clW)}$ as
  fibrations.
\end{theorem}

\begin{theorem}
  There is a model structure on the category~$\rPres$ of reflexive presentations
  with~$\clW$ (as defined in \cref{sec:pres-w}) as weak equivalences,
  $\clC=\lrlp\clI$ as cofibrations (with $\clI$ as defined in
  \cref{sec:pres-gen-cof}) and $\clF=\rlp{(\clC\cap\clW)}$ as fibrations.
\end{theorem}
\begin{proof}
  We apply \cref{thm:smith}. First point was shown in \cref{lem:pres-w-23},
  second point in \cref{lem:pres-tfib-w}. For third point, the closure of
  $\lrlp\clI$ under pushouts and countable compositions was shown in
  \cref{lem:ortho} and the one of $\clW$ in \cref{lem:pres-w-23}, from which we
  deduce the one of their intersection. The last point is the object of
  \cref{prop:solution-set}.
\end{proof}

\subsection{A Quillen functor}
The category $\Mon$ can canonically be equipped with the \emph{trivial model
  structure} where weak equivalences are isomorphisms and every morphism is both
fibrant and cofibrant. The presentation functor $\rPres\to\Mon$ described in
\secr{pmon} is a left adjoint (\lemr{pmon-adj} and \remr{refl-stdpres}) which
trivially preserves cofibrations and trivial cofibrations, and is thus a Quillen
functor. Moreover, this functor reflects weak equivalences and, given a
presentation~$\P$, the counit $\P\to\stdpres{\pmon\P}$ of the adjunction is a
weak equivalence: by \cite[Corollary~1.3.16]{hovey2007model}, the presentation
functor is thus a Quillen equivalence. By
\cite[Proposition~1.3.13]{hovey2007model}, this means that the derived functor
induces, as expected, an equivalence of categories between the localization of
$\rPres$ under weak equivalences and the one of~$\Mon$ (which is $\Mon$ itself):
\[
  \Ho(\rPres)
  \equivto
  \Ho(\Mon)
  \isoto
  \Mon
  \pbox.
\]
\section{Tietze transformations as trivial cofibrations}
\label{sec:tietze-tcof}
In \secr{clJ} below, we introduce a class~$\clJ$ of morphisms of reflexive
presentations such that pushouts of morphisms in this class correspond to
elementary Tietze transformations. Contrarily to what one could expect, this
family does not generate all trivial cofibrations: we have a strict inclusion
$\lrlp\clJ\subsetneq\clC\cap\clW$. However, we show that the two classes
coincide for morphisms with fibrant codomain: we thus say that the class~$\clJ$
is pseudo-generating, following the terminology of
Simpson~\cite[Section~8.7]{simpson2011homotopy}.

\subsection{Pseudo-generating trivial cofibrations}
\label{sec:clJ}
We write $\clJ$ for the class of morphisms of $\rPres$, called
\emph{pseudo-generating trivial cofibrations}
\begin{align*}
  \pres{a_1,\ldots,a_m}{}&\into\pres{a_1,\ldots,a_m,a_{m+1}}{u\rto a_{m+1}}\\
  \pres{a_1,\ldots,a_m}{}&\into\pres{a_1,\ldots,a_m}{u\rto u}\\
  \pres{a_1,\ldots,a_{m+n}}{u\rto v}&\into\pres{a_1,\ldots,a_{n+m}}{u\rto v,v\rto u}\\
  \pres{a_1,\ldots,a_{m+n+p}}{u\rto v,v\rto w}&\into\pres{a_1,\ldots,a_{n+m+p}}{u\rto v,v\rto w,u\rto w}\\
  \pres{a_1,\ldots,a_{m+n+p+q}}{u\rto v}&\into\pres{a_1,\ldots,a_{m+n+p+q}}{wuw'\rto wvw'}
\end{align*}
for some $m,n,p\in\N$ with
\begin{align*}
  u&=a_1\ldots a_m
  &
  w&=a_{m+n+1}\ldots a_{m+n+p}
  \\
  v&=a_{m+1}\ldots a_{m+n}
  &
  w'&=a_{m+n+p+1}\ldots a_{m+n+p+q}
\end{align*}

\begin{lemma}
  \label{lem:gtcof-tietze}
  Given a pseudo-generating cofibrations $j:\P\to\Q$ and a morphism of
  presentations $f:\P\to\P'$, consider the pushout $j':\P'\to\Q'$ of $j$
  along~$f$:
  \[
    \begin{tikzcd}
      \P\ar[d,"j"']\ar[r,"f"]&\P'\ar[d,dotted,"j'"]\\
      \Q\ar[r,dotted]&\Q'
    \end{tikzcd}
  \]
  then there is an elementary Tietze transformation from~$\P'$ to~$\Q'$, and
  conversely every elementary Tietze transformation arises in this way.
\end{lemma}
\begin{proof}
  Pushout of the five kinds of morphisms in~$\clJ$ precisely give rise to the
  five kinds of Tietze transformations \tgen{}, \trefl{}, \tsym{}, \ttrans{} and
  \tctxt{}.
\end{proof}

\noindent
We are thus tempted to call \emph{generalized Tietze transformation} a morphism
in~$\cell\clJ$.
In particular, every element of~$\clJ$ is itself a Tietze transformation and
thus, by \thmr{tietze},

\begin{lemma}
  \label{lem:gtcof-w}
  Generating trivial cofibrations are weak equivalences: $\clJ\subseteq\clW$.
\end{lemma}

\noindent
Moreover, those morphisms are monomorphisms and thus, by \lemr{mon-cof},

\begin{lemma}
  \label{lem:gtcof-cof}
  The pseudo-generating trivial cofibrations are cofibrations:
  $\clJ\subseteq\lrlp\clI$.
\end{lemma}

\begin{remark}
  By general properties~\cite[Proposition~2.1.18]{hovey2007model}, we have that
  morphisms in $\lrlp\clJ$ are retracts of Tietze transformations. We do not
  know whether the morphisms in $\lrlp\clJ$ are precisely Tietze transformations
  or not.
\end{remark}

\subsection{Morphisms in~$\lrlp\clJ$}
The following lemmas show that the morphisms in the class $\lrlp\clJ$ are
trivial cofibrations. We will however see in \secr{pseudo-not-gen} that not
every trivial cofibration is in this class, \ie the inclusion is strict.

\begin{lemma}
  \label{lem:pres-tcof-cof}
  We have $\lrlp\clJ\subseteq\lrlp\clI$.
\end{lemma}
\begin{proof}
  By \lemr{gtcof-cof}, we have that $\clJ\subseteq\lrlp\clI$. Thus, by
  \lemr{ortho}, we have
  \[
    \llp{(\rlp\clJ)}\subseteq\lrlp{(\lrlp\clI)}=\lrlp\clI
    \pbox.
    \qedhere
  \]
\end{proof}

\begin{lemma}
  \label{lem:pres-tcof-w}
  We have $\lrlp\clJ\subseteq\clW$.
\end{lemma}
\begin{proof}
  %
  By \lemr{gtcof-tietze}, a pushout of an element in~$\clJ$ is an elementary
  Tietze transformation and thus a weak equivalence by \lemr{tietze-iso}. By
  \propr{small-obj}, any element of $\lrlp\clJ$ is a countable composition of
  elementary Tietze transformations, and thus a weak equivalence by
  \lemr{pres-w-comp}.
\end{proof}

\begin{lemma}
  \label{lem:pres-tcof-tcof}
  We have $\lrlp\clJ\subseteq\clC\cap\clW$.
\end{lemma}
\begin{proof}
  By Lemmas~\ref{lem:pres-tcof-cof} and \ref{lem:pres-tcof-w}.
\end{proof}

\subsection{Pseudo-fibrations}
\label{sec:pres-fib}
The morphisms in $\rlp\clJ$ are called \emph{pseudo-fibrations}. A
\emph{pseudo-fibrant object}~$\P$ is one such that the terminal morphism
$\P\to\tobj$ is a pseudo-fibration.

\begin{lemma}
  \label{lem:pfib-obj}
  A presentation~$\P$ is pseudo-fibrant when
  \begin{itemize}
  \item for every word $u\in\freemon\P_1$, there is a generator $a\in\P_1$ such
    that $u\rto a\in\P_2$,
  \item the relation $\P_2$ on $\freemon\P_1$ is a congruence.
  \end{itemize}
  In particular, we have $u\psim\P v$ if and only if $u\rto v\in\P_2$.
\end{lemma}

\noindent
More generally, pseudo-fibrations can be described as follows:

\begin{lemma}
  \label{lem:pfib}
  A morphism $f:\P\to\Q$ is a pseudo-fibration when
  \begin{itemize}
  \item for every $u\in\freemon\P_1$ and $b\in\Q_1$ such that
    $f(u)\rto b\in\Q_2$, there exists $a\in\P_1$ with $\freemon f(a)=b$ and
    $u\rto a\in\P_2$,
  \item for every $u\in\freemon\P_1$,
    \[
      \freemon f(u)\rto \freemon f(u)\in\Q_2
      \qqtimpl
      u\rto u\in\P_2
      \pbox,
    \]
  \item for every $u,v\in\freemon\P_1$ with $u\rto v\in\P_2$,
    \[
      \freemon f(v)\rto \freemon f(u)\in\Q_2
      \qqtimpl
      v\rto u\in\P_2
      \pbox,
    \]
  \item for every $u,v,w\in\freemon\P_1$ with $u\rto v\in\P_2$ and
    $v\rto w\in\P_2$,
    \[
      \freemon f(u)\rto \freemon f(w)\in\Q_2
      \qqtimpl
      u\rto w\in\P_2
      \pbox,
    \]
  \item for every $u,v,w,w'\in\freemon\P_1$ with $u\rto v\in\P_2$,
    \[
      \freemon f(wuw')\rto\freemon f(wvw')\in Q_2
      \qqtimpl
      wuw'\rto wvw'\in\P_2
      \pbox.
    \]
  \end{itemize}
\end{lemma}

\begin{lemma}
  \label{lem:fib-pfib}
  Any fibration is a pseudo-fibration:
  $\clF=\rlp{(\clC\cap\clW)}\subseteq\rlp\clJ$.
\end{lemma}
\begin{proof}
  By \lemr{pres-tcof-tcof}, we have $\lrlp\clJ\subseteq\clC\cap W$. Therefore,
  by \lemr{ortho},
  \[
    \clF=\rlp{(\clC\cap\clW)}\subseteq\rlp{(\lrlp\clJ)}=\rlp\clJ
    \pbox.
    \qedhere
  \]
\end{proof}

\begin{lemma}
  \label{lem:pfib-repl}
  For any object~$\P$, there exists a pseudo-fibrant object $\tilde\P$, called a
  \emph{pseudo-fibrant replacement} of~$\P$, together with a map~$\P\to\tilde\P$
  in $\lrlp\clJ$.
\end{lemma}
\begin{proof}
  Use the small object argument (\propr{small-obj}) to factor the terminal
  morphism $\P\to\tobj$ as a morphism in $\lrlp\clJ$ followed by a morphism in
  $\rlp\clJ$.
\end{proof}

\subsection{$\clJ$ is not generating}
\label{sec:pseudo-not-gen}
Contrarily to what one might expect, the class $\clJ$ is not a generating class for
the trivial cofibrations. This can be seen by observing that the following inclusion does
not hold:
\[
  \rlp\clJ\cap\clW\subseteq\rlp\clI
\]
For instance, consider the inclusion
\[
  \pres{a}{}\to\pres{a,b}{b\rto bb,1\rto bb}
\]
which corresponds to the example developed~\secr{ex-tietze}. This morphism is
both a pseudo-fibration since the only relations to lift are the reflexivity
relations (which are not noted here, see \secr{refl-pres}) and a weak
equivalence since both presented monoids are~$\N$. However, it is not a trivial
fibration since it is not surjective on generators.
The same example can be used to show that the inclusion
\[
  \lrlp\clI\cap\clW\subseteq\lrlp\clJ
\]
does not hold either: the map above is a trivial cofibration since it is both a
monomorphism and a weak equivalence, but it cannot be obtained as a retract
of a composite of pushouts of sums of elements of~$\clJ$. Namely, the
generator~$b$ has to be added using a Tietze transformation \tgen{}, but the
relations are not of the right form. Intuitively, the relation $1\rto b$ has to
be added first, see \secr{ex-tietze}.

\begin{remark}
  As a simpler (but less illuminating) example, consider the inclusion
  \[
    \pres{a}{}\to\pres{a,b}{b\rto aa}
  \]
  which is not an elementary Tietze transformation, because of the chosen
  orientation for the relation~\tgen{}.
  %
  Similarly, the inclusion
  \[
    \pres{a,b,c,d}{aa\rto bb,bb\rto cc,cc\rto dd}
    \to
    \pres{a,b,c,d}{aa\rto bb,bb\rto cc,cc\rto dd,aa\rto dd}
  \]
  is a pseudo-fibration and a weak equivalence, but not a trivial fibration one
  since the relation $aa\rto dd$ cannot be lifted.
\end{remark}

\subsection{$\clJ$ is pseudo-generating}
It is interesting to note that the inclusions of previous section are satisfied
if we restrict to fibrations whose codomain is fibrant. We begin by a reciprocal
to \lemr{pres-tcof-tcof}:

\begin{lemma}
  \label{lem:tcof-tcof}
  Any trivial cofibration $i:\P\to\Q$ with pseudo-fibrant codomain~$\Q$ belongs
  to $\cell\clJ$, and thus to $\lrlp\clJ$.
\end{lemma}
\begin{proof}
  Since $i$ is a trivial cofibration, it is an injection and we have
  $\pmon\P=\pmon\Q$. For simplicity, we suppose that $i$ is an inclusion.
  For every generator in $a\in\Q_1\setminus\P_1$, there is a word
  $u_a\in\freemon\P_1$ such that $u_a\psim\Q a$ and therefore $u_a\rto a\in\Q_2$
  since $\Q$ is pseudo-fibrant ($\Q_2$ is a congruence). Writing $\P^0$ for~$\P$
  with the generator $a$ and a relation $u_a\rto a$ added, for every
  $a\in\Q_1\setminus\P_1$, we have a morphism $\P\to\P^0$ in $\cell\clJ$
  factoring~$f$ (the inclusion $\P\to\P^0$ can be expressed as a pushout of a
  coproduct of pseudo-generating trivial cofibrations of the first form). We
  write $\P^{k+1}$ for the presentation obtained from~$\P^k$ by adding
  \begin{itemize}
  \item a relation $u\rto u$ for every word $u$ over $\P^k_1$,
  \item a relation $v\rto u$ for every relation $u\rto v\in P^k_2$,
  \item a relation $u\rto w$ for every relations $u\rto v,v\rto w\in P^k_2$,
  \item a relation $wuw'\rto wvw'$ for every relation $u\rto v\in P^k_2$ and
    words $w,w'$ over $\P^k_1$.
  \end{itemize}
  There is a morphism $\P^k\to\P^{k+1}$ in $\cell\clJ$. Every generator of~$\Q$
  gets added at the first step and every relation of~$\Q$ gets added at some
  step.
  Therefore $\Q=\colim_k\P^k$ and $i$ belongs to $\cell\clJ$.
\end{proof}

\begin{remark}
  The above proof essentially consists in using the small object argument to
  construct a factorization $i=g\circ f$ with $f\in\cell\clJ$ and
  $g\in\rlp\clJ$, and observing that $g$ can be chosen to be an identity when
  $\Q$ is pseudo-fibrant.
\end{remark}

\begin{lemma}
  \label{lem:pres-w-cof-fib}
  Any pseudo-fibration $p:\P\to\Q\in\rlp\clJ$ with pseudo-fibrant target~$\Q$ is
  a fibration, \ie $p\in\rlp{(\clC\cap\clW)}$.
\end{lemma}
\begin{proof}
  Suppose given a trivial cofibration $i:\P'\to\Q'\in\clC\cap\clW$ and two
  morphisms $f:\P'\to\P$ and $g:\Q'\to\Q'$ such that $p\circ f=g\circ i$. By
  \lemr{pfib-repl}, we can consider a pseudo-fibrant replacement $\tilde\Q'$
  of~$\Q'$ together with the associated morphism $j:\Q'\to\tilde\Q'$ in
  $\lrlp\clJ$, and thus in $\clC\cap\clW$ by \lemr{pres-tcof-tcof}. By
  orthogonality, there is a map $k:\tilde\Q'\to\Q$ such that $k\circ j=g$.
  Finally, by \lemr{tcof-tcof} $(j\circ i)\boxslash p$, from which we deduce the
  existence of $h:\tilde\Q'\to\P$ such that $h\circ j\circ i=f$ and
  $p\circ h=k$.
  \[
    \begin{tikzcd}
      \P'\ar[d,"\clC\cap\clW\owns i"']\ar[r,"f"]&\P\ar[d,"p\in\rlp\clJ"]\\
      \Q'\ar[d,"\clC\cap\clW\supseteq\lrlp\clJ\owns j"']\ar[r,"g"']&\Q\ar[d,"\in\rlp\clJ"]\\
      \tilde\Q'\ar[r,"\in\rlp\clJ"']\ar[ur,"k"description]\ar[uur,"h"description,near end,bend left=10]&\tobj
    \end{tikzcd}
  \]
  Therefore the morphism $h\circ j:\Q'\to\P$ is a filler and thus
  $i\boxslash p$.
\end{proof}

\begin{lemma}
  \label{lem:pfib-fib-obj}
  Pseudo-fibrant and fibrant objects coincide.
\end{lemma}
\begin{proof}
  By \lemr{fib-pfib}, any fibrant object is pseudo-fibrant. Conversely, by
  \lemr{pres-w-cof-fib}, it suffices to check that the terminal object is
  pseudo-fibrant, which can be verified directly.
\end{proof}

\begin{lemma}
  Given a monoid~$M$, its standard presentation~$\stdpres{M}$ is fibrant.
\end{lemma}
\begin{proof}
  The presentation~$\stdpres{M}$ satisfies the conditions of \lemr{pfib-obj} and
  is thus pseudo-fibrant and thus fibrant by \lemr{pfib-fib-obj}.
\end{proof}

\section{Tietze equivalences as cospans}
\label{sec:teq}
In this section we reconstruct the proof of the Tietze theorem by showing that
any two presentations of the same monoid can be related by a cospan of
generalized Tietze transformations.

\subsection{Coproduct}
We begin by showing that, under suitable hypothesis, the canonical injections
into coproducts are cofibrations.

\begin{lemma}
  \label{lem:inj-cof}
  In a model category, when~$X$ is cofibrant, the canonical injections
  $\iota_0:Y\to Y\sqcup X$ and $\iota_1:Y\to X\sqcup Y$ are cofibrations.
\end{lemma}
\begin{proof}
  We have a pushout diagram
  \[
    \begin{tikzcd}
      \iobj\ar[d]\ar[r]\ar[phantom,dr,"\ulcorner",very near end]&Y\ar[d,"\iota_1"]\\
      X\ar[r,"\iota_0"']&X\sqcup Y
    \end{tikzcd}
  \]
  When~$X$ is cofibrant, the initial map into~$X$ is a cofibration, and the
  map~$\iota_1$ is thus also a cofibration, as a pushout of a cofibration. The
  other case is similar.
\end{proof}

\subsection{Weak equivalences as cospans}
We now recall the contents of the proof of the celebrated Ken Brown lemma, which
shows that every weak equivalence between cofibrant objects factors as a cospan
of trivial cofibrations.

\begin{lemma}[Ken Brown's lemma]
  \label{lem:ken-brown}
  In a model category, every weak equivalence $w:X\to Y$ between cofibrant
  objects~$X$ and~$Y$ factors as $w=p\circ i$ where $i$ is a trivial cofibration
  and $p$ a trivial fibration which admits a section by a trivial
  cofibration~$j$:
  \[
    \begin{tikzcd}
      &Z\ar[dr,->>,bend left,"p"]&\\
      X\ar[ur,hook,"i"]\ar[rr,"w"']&&\ar[ul,hook',"j"]Y\pbox.
    \end{tikzcd}
  \]
\end{lemma}
\begin{proof}
  We can factor the map $(w,\id_Y):X\sqcup Y\to Y$ as a cofibration
  $k:X\sqcup Y\to Z$ followed by a trivial fibration $p:Z\to Y$. Since~$X$
  and~$Y$ are cofibrant, by \lemr{inj-cof}, the injections into $X\sqcup Y$ are
  cofibrations. We define $i=k\circ\iota_0$ and $j=k\circ\iota_1$:
  \[
    \begin{tikzcd}[row sep=small]
      X\ar[dr,hook,"\iota_0"]\ar[drr,bend left,"i"description]\ar[drrr,bend left,"w"]\\
      &X\sqcup Y\ar[r,hook,"k"]&Z\ar[r,->>,"p"]&Y\\
      Y\ar[ur,hook,"\iota_1"']\ar[urr,bend right,"j"description]\ar[urrr,bend right,"\id_Y"']
    \end{tikzcd}
  \]
  The maps $i$ and $j$ are cofibrations as composites of cofibrations and are
  weak equivalences by the 2-out-of-3 property.
\end{proof}

\begin{remark}
  In the previous lemma, the cospan $(i,j)$ can be considered as a factorization
  of $w$, in the sense that we have $j\circ w=j\circ p\circ i=i$.
\end{remark}

\begin{remark}
  In a model category where monomorphisms are cofibrations (such as the case of
  interest here, see \lemr{cof-mon}), a simpler argument can be given: since~$Y$
  is cofibrant and $p$ is a trivial fibration, the diagram
  \[
    \begin{tikzcd}
      \iobj\ar[d,hook]\ar[r]&Z\ar[d,->>,"p"]\\
      Y\ar[r,"\id_Y"']\ar[ur,dotted,"j"description]&Y
    \end{tikzcd}
  \]
  admits a filler~$j:Y\to Z$, which is a section of~$p$; moreover, since $j$ is
  a monomorphism, it is a cofibration, and it is a weak equivalence by the
  2-out-of-3 property.
\end{remark}

\begin{theorem}
  In a model category~$\M$ in which every object is cofibrant, every isomorphism
  in~$\Ho(\M)$ is the localization of a cospan of trivial cofibrations.
\end{theorem}
\begin{proof}
  Consider an isomorphism~$f:X\to Y$ in~$\Ho(\M)$. We write $\M'$ for the full
  subcategory of~$\M$ whose objects are fibrant. The fibrant replacement
  functor~$F:\M\to\M'$ induces an equivalence between the homotopy
  categories~\cite[Proposition~1.2.3]{hovey2007model}. Moreover, $\Ho(\M')$ is a
  quotient of~$\M'$ by homotopy
  equivalences~\cite[Theorem~1.2.10]{hovey2007model}, the map $Ff$ is thus a
  homotopy equivalence and thus a weak
  equivalence~\cite[Proposition~1.2.8]{hovey2007model}. The map~$f$ is thus the
  localization of a span of weak equivalences
  \[
    \begin{tikzcd}
      X\ar[d,hook,"i_X"']&Y\ar[d,hook,"i_Y"]\\
      FX\ar[r,"Ff"']&FY
    \end{tikzcd}
  \]
  where $i_X:X\to FX$ is the trivial cofibration associated to the fibrant
  replacement. By \lemr{ken-brown}, we thus have two cospans of trivial
  cofibrations
  \[
    \begin{tikzcd}[sep=small]
      &X'&&Y'\\
      X\ar[ur,hook]&&\ar[ul,hook']FY\ar[ur,hook]&&\ar[ul,hook']Y\\
    \end{tikzcd}
  \]
  and we conclude to the existence of one cospan of trivial cofibrations using
  the fact that trivial cofibrations are closed under pushouts.
\end{proof}

\subsection{Tietze equivalences}
We can now conclude with the abstract proof of the Tietze theorem.

\begin{theorem}
  \label{thm:gen-tietze}
  In the category $\rPres$, two presentations~$\P$ and~$\Q$ are such that
  $\pmon\P\isoto\pmon\Q$ if and only if there is a cospan of generalized Tietze
  transformations (of morphisms in $\cell\clJ$) from~$\P$ to~$\Q$.
\end{theorem}
\begin{proof}
  Suppose given two presentations~$\P,\Q\in\rPres$ such that
  $\pmon\P\isoto\pmon\Q$. With the model structure introduced in
  \secr{rpres-mod}, this can be rewritten as $\Ho(\P)\isoto\Ho(\Q)$, and
  therefore we deduce that there is a cospan of trivial cofibrations
  \[
    \begin{tikzcd}[sep=small]
      &\R\\
      \P\ar[ur,hook]&&\ar[ul,hook']\Q\pbox.
    \end{tikzcd}
  \]
  Up to taking a fibrant replacement of~$\R$ and suppose that~$\R$ is fibrant
  and thus pseudo-fibrant by \lemr{pfib-fib-obj}. We deduce that this is a span
  of Tietze transformations by \lemr{tcof-tcof}.
  Conversely, Tietze transformations are weak equivalences by~\lemr{pres-tcof-w}
  and thus~$\P$ and~$\Q$ become isomorphic after localizing under weak
  equivalences.
\end{proof}

\section{Variants and extensions}
Many variants of the situation considered here could be thought of and are left
for future work.

\subsection{Non-reflexive presentations}
\label{sec:nr-pres}
If we consider the category~$\Pres$ of (non-necessarily reflexive)
presentations, many of the constructions performed in previous section can still
be carried over. However, \lemr{pres-tfib-w} does not hold anymore, preventing
the construction of a model category: the elements of $\rlp\clI$ are not
necessarily weak equivalences. As a counter-example consider the morphism
\[
  \pres{a,b}{}\to\pres{c}{}
  \pbox.
\]
It belongs to~$\rlp\clI$ since it satisfies the conditions of \lemr{tfib} (which
still holds): it is surjective on generators and lifts every required relation
since there are none. It is however not a weak equivalence since the monoids
presented by the source and the target are respectively $\N*\N$ and $\N$ which
are not isomorphic (the first one is not commutative for instance). We expect
that there is however a right semi-model structure in the sense
of~\cite{barwick2010left},
whose cofibrations are generated by~$\clI$.

\subsection{Multisets of relations}
\label{sec:pres-mon-pol}
The notion of presentation can be modified in order to allow multiple relations
with the same source and the same target: such a presentation~$\P$ consists of a
set~$\P_1$ of generators together with a set~$\P_2$ of relations equipped with
source and target maps $\src,\tgt:\P_2\to\P_1$. Here, an element $\alpha\in\P_2$
with $\src(\alpha)=u$ and $\tgt(\alpha)=v$ encodes a relation $u\rto v$. We
expect that this modification does not significantly changes the situation
studied here.

\subsection{Presentations of categories}
\label{sec:pres-pol}
As a further generalization, one can consider presentations of categories. Such
a presentation~$\P$ of a category consists of a set~$\P_0$ of objects, a
set~$\P_1$ of generators for morphisms equipped with source and target maps
$\src_0,\tgt_0:\P_1\to\P_0$, and a set~$\P_2$ of relations equipped with source
and target maps $\src_1,\tgt_1:\P_2\to\P_1^*$ such that
$\src_0^*\circ\src_1=\src_0^*\circ\tgt_1$ and
$\tgt_0^*\circ\src_1=\tgt_0^*\circ\tgt_1$. Here, $\P_1^*$ denotes the morphisms
of the free category over the graph $(\P_0,\P_1)$ and the category presented
by~$\P$ is obtained by quotienting the morphisms of this free category under the
congruence generated by~$\P_2$. The notion of presentation of monoid of
\secr{pres-mon-pol}, is the particular case where $\P_0=\set{\star}$ is reduced
to one element. We expect the proofs of this paper to generalize to this setting.

\subsection{Presentations of $n$-categories}
This notion of presentation sketched in the previous section, is a particular case
of the notion of \emph{polygraph}, see~\cite{burroni1993higher}, which
generalizes to give presentations of $n$-categories. It would be interesting to see whether
the model structure extends to this case.

\subsection{Presentations of groupoids}
The notion of Tietze transformation was originally developed for presentations
of groups. It would be interesting to generalize the model structure to this
case, as well as generalizations of presentations of groupoids.

\subsection{Coherent presentations}
A notion of Tietze transformation for coherent presentations of categories is
introduced in~\cite{gaussent2015coherent}. We would like to investigate this
case, as well as, more generally, developing a notion of Tietze transformation for
resolutions of categories by $(\infty,1)$-polygraphs.


\bibliographystyle{plain}
\bibliography{article}
\end{document}